\documentclass[fleqn,10pt]{wlscirep}

\usepackage[utf8]{inputenc}
\usepackage[T1]{fontenc}
\usepackage{amsmath}
\usepackage{amssymb}
\usepackage{mathtools}
\usepackage{upgreek}
\usepackage{cite}

\title{Energy-optimized scaling laws for self-guided laser wakefield accelerators}

\author[1,*]{Petr Valenta}
\author[1]{Marcel Lamač}
\author[2]{Kyle G. Miller}
\author[3]{Brandon K. Russell}
\author[1]{Gabriele~M.~Grittani}
\author[4]{Alec G. R. Thomas}
\author[1]{Sergei V. Bulanov}

\affil[1]{ELI Beamlines Facility, The Extreme Light Infrastructure ERIC, Dolní Břežany, Czechia}
\affil[2]{Laboratory for Laser Energetics, University of Rochester, Rochester, NY, USA}
\affil[3]{Department of Astrophysical Sciences, Princeton University, Princeton, NJ, USA}
\affil[4]{Gerard Mourou Center for Ultrafast Optical Science, University of Michigan, Ann Arbor, MI, USA}

\affil[*]{petr.valenta@eli-laser.eu}

\keywords{laser wakefield acceleration, self-guiding, scaling laws, particle-in-cell simulation, Bayesian optimization}

\begin{abstract}
Laser wakefield acceleration promises compact electron accelerators for applications in medicine, industry, and fundamental science. Yet, despite rapid progress, accurately predicting the electron energy attainable in a given experimental configuration and the acceleration length required to reach it remains an open challenge. Here we use Bayesian optimization combined with advanced particle-in-cell simulation techniques to determine the maximum electron energy that a self-guided laser wakefield accelerator driven by a laser of a given energy and wavelength can produce. By systematically optimizing the accelerator performance across a range of laser energies and wavelengths, we derive energy-optimized scaling laws. These scaling laws yield the highest electron energy over the shortest acceleration length possible, are expressed solely in terms of laser energy and wavelength, and are accompanied by the complete set of laser and plasma parameters required to enable the scaling. The resulting scaling laws provide practical guidance for designing state-of-the-art laser wakefield acceleration experiments operating at their fundamental performance limits.
\end{abstract}

\begin{document}

\flushbottom

\maketitle

\thispagestyle{empty}

\section*{Introduction}

Laser wakefield acceleration (LWFA) has emerged as a powerful electron acceleration scheme in which an intense laser pulse excites a plasma wave that traps and accelerates electrons to relativistic energies~\cite{tajima1979, malka2008, esarey2009, bulanov2016, kato2026}. Owing to its ability to sustain accelerating fields several orders of magnitude stronger than those of conventional radio-frequency accelerators, LWFA has the potential to dramatically reduce the footprint of electron accelerators. This capability, together with the prospect of delivering electron beams with GeV-scale energies~\cite{gonsalves2019, aniculaesei2023, picksley2024}, nC-level charges~\cite{couperus2017, rockafellow2025}, and kHz repetition rates~\cite{guenot2017, faure2018, salehi2021, lazzarini2024}, has generated considerable interest in LWFA as a compact accelerator platform for a wide range of applications, including the production of photons~\cite{gruner2007, corde2013, bulanov2013, albert2014, kurz2021}, positrons~\cite{williams2015, streeter2024}, and muons~\cite{zhang2025, ludwig2025, terzani2025, calvin2026}. Furthermore, the co-location of LWFA with multi-petawatt laser facilities offers unique opportunities to explore previously inaccessible regimes of non-perturbative strong-field quantum electrodynamics~\cite{yan2017, cole2018, poder2018, gonoskov2022, russell2023, mirzaie2024, russell2024}.

Despite tremendous progress, many aspects of the fundamental physics governing LWFA remain an active area of research, particularly with regard to the reliable control and optimization of the acceleration process. The underlying challenge stems from the inherently complex nature of LWFA, which simultaneously involves multiphysics, multiscale, and multiparametric behavior. Analytical models provide valuable physical insight but often rely on simplifying assumptions that limit their predictive capability under realistic experimental conditions. Likewise, experiments are constrained by practical considerations such as available laser technology, target reproducibility, and diagnostic resolution. As a result, progress has often relied on localized optimization within specific operating regimes, which restricts the broader applicability of the conclusions.

To overcome these limitations, increasing emphasis has been placed on advanced numerical and optimization strategies. Contemporary high-performance computing platforms enable detailed simulations that faithfully reproduce the dominant physics at manageable computational cost. Nevertheless, exhaustively scanning the vast, high-dimensional parameter space remains impractical. In this context, artificial intelligence and machine-learning techniques have emerged as effective tools for guiding exploration, locating optimal regimes, and significantly reducing the overall computational burden~\cite{shalloo2020, jalas2021, kirchen2021, dopp2023a, roussel2024, irshad2024, valenta2025, valenta2026, maslarova2026}.

From a physics perspective, efficient LWFA requires that the driving laser pulse maintains a high intensity over distances far exceeding its diffraction length. This necessitates some form of guiding within the plasma, which can be provided either externally or through self-guiding. In this work, we concentrate on the latter. Self-guided propagation is appealing from an experimental standpoint because it avoids the need for precisely engineered guiding structures. On the other hand, self-guiding typically occurs in a strongly nonlinear regime, which can compromise stability and beam quality compared with externally guided configurations~\cite{esarey2009, picksley2024}. These considerations make self-guided LWFA an attractive yet nontrivial candidate for systematic optimization.

The onset of self-guiding is governed primarily by the laser power, $ \mathcal{P}_0 $, relative to the threshold for relativistic self-focusing,
\begin{equation}\label{eq:1}
    \mathcal{P}_{\mathrm{cr}} = \overline{\mathcal{P}} \frac{n_{\mathrm{cr}}}{n_{\mathrm{e}}},
\end{equation}
where $ n_{\mathrm{e}} $ is the electron density, $ n_{\mathrm{cr}} = \uppi / r_{\mathrm{e}} \lambda_0^2 $ is the critical plasma density corresponding to the laser wavelength $ \lambda_0 $, and $ \overline{\mathcal{P}} = 2 m_{\mathrm{e}} c^3 / r_{\mathrm{e}} \approx 17.4~\mathrm{GW} $~\cite{sun1987}. Here, $ m_{\mathrm{e}} $ is the electron mass, $ r_{\mathrm{e}} = e^2 / 4 \uppi \epsilon_0 m_{\mathrm{e}} c^2 $ is the classical electron radius, $ c $ is the speed of light in vacuum, $ \epsilon_0 $ is the vacuum permittivity, and $ e $ is the elementary charge. When $ \mathcal{P}_0 < \mathcal{P}_{\mathrm{cr}} $, diffraction dominates, causing the laser beam to expand and its intensity to decrease, thereby reducing the acceleration length and the attainable electron energy. Conversely, when $ \mathcal{P}_0 \gg \mathcal{P}_{\mathrm{cr}} $, strong nonlinear effects (e.g., laser filamentation~\cite{naumova2002, valenta2021}, vortex generation~\cite{bulanov1996}, and soliton formation~\cite{bulanov1999, esirkepov2002}) can distort the beam profile and degrade the acceleration process. Although high electron energies may still be achieved in this strongly nonlinear regime, here we focus on stable, efficient, and reproducible self-guided LWFA, which requires the laser power to remain within an appropriate range relative to the critical power.

In addition, stable self-guided laser pulse propagation in plasma requires the so-called matching conditions~\cite{lu2006,lu2007}
\begin{equation}\label{eq:2}
    k_{\mathrm{p}} w_0 \approx 2 \sqrt{a_0} \qquad \textnormal{and} \qquad
    a_0 \approx 2 \left( \frac{\mathcal{P}_0}{\mathcal{P}_{\mathrm{cr}}} \right)^{1/3},
\end{equation}
where $ w_0 $ is the laser waist, $ a_0 = e E_0 / m_{\mathrm{e}} \omega_0 c $ is the normalized laser amplitude, and $ k_{\mathrm{p}} = \omega_{\mathrm{p}} / c $ is the plasma wavenumber, with $ \omega_0 $ and $ \omega_{\mathrm{p}} $ denoting the laser and plasma frequencies, respectively, and $ E_0 $ the laser electric-field amplitude. These conditions are obtained by balancing the laser ponderomotive force against the plasma restoring force and express the requirement that the laser beam waist be comparable to the radius of the ion cavity formed behind the pulse. The formulation of the matching conditions in Eqs.~\eqref{eq:2} was verified to be optimal for maximizing electron energy in self-guided LWFA, and the allowed variation ranges of the parameters entering these conditions for maintaining high-energy operation were estimated~\cite{valenta2026}.

Classical scaling relations for the maximum electron energy and the corresponding acceleration length in self-guided LWFA were established two decades ago~\cite{kostyukov2004, gordienko2005, lu2006, lu2007}, and have since been validated, refined, and extended in a variety of operating regimes and optimization scenarios~\cite{tzoufras2009, schroeder2010, yi2013, bulanov2016, davidson2019, dalichaouch2021, golovanov2023}. Building on these seminal results, in the present work we reformulate the scaling laws such that they (i) yield the highest electron energy over the shortest acceleration length possible, (ii) are expressed solely in terms of the laser energy and wavelength, and (iii) are accompanied by the complete set of laser and plasma parameters required to enable the scaling. The resulting scaling laws are directly applicable to the design of LWFA experiments operating at their fundamental performance limits and emphasize the fact that laser energy and wavelength—not peak power—are the natural design parameters for LWFA facilities.

To accomplish our goal, we employ Bayesian optimization (BO), a machine-learning method well suited to optimizing objective functions that are expensive to evaluate. In our case, the objective function is the maximum electron energy, and its values are sampled using particle-in-cell (PIC) simulations performed in a quasi-three-dimensional (quasi-3D) geometry~\cite{lifschitz2009, davidson2015} and in a Lorentz-boosted frame~\cite{vay2007, yu2016}. At the current stage, we concentrate on the maximum electron energy alone, leaving aside other beam characteristics (e.g., charge, energy spread, and divergence). This choice is motivated by applications where surpassing a certain energy threshold is the primary requirement, such as muon generation~\cite{zhang2025, ludwig2025, terzani2025, calvin2026} or nuclear activation~\cite{nedorezov2021, kolenaty2022}. Furthermore, while electron energy is determined mainly by the parameters of the laser and the background plasma, beam quality metrics are expected to depend strongly on a particular electron injection mechanism. Therefore, the energy-optimized scaling laws identified here remain applicable across various injection schemes, with beam-quality optimization addressed separately.

To reduce the dimensionality of the optimization problem, we impose the following assumptions: (i) LWFA is driven by a linearly polarized Gaussian laser pulse with energy $ \mathcal{E}_0 $, wavelength $ \lambda_0 $, and constant spectral phase; (ii) the plasma has a uniform electron density $ n_{\mathrm{e}} $; (iii) acceleration occurs in a single-stage, self-guided, matched regime satisfying Eqs.~\eqref{eq:2}; and (iv) electrons are injected externally and treated as test particles. These assumptions isolate the dominant physical dependencies while keeping the computational effort tractable. Under these assumptions, however, the identified maximum electron energy does not necessarily represent the global maximum. More complex configurations, such as LWFA driven by non-Gaussian~\cite{beaurepaire2015, oumbarekespinos2023} or frequency-chirped~\cite{kalmykov2012, kim2017a} laser pulses, as well as using tailored plasma profiles~\cite{bulanov1993, bulanov1997, sprangle2001, guillaume2015, ludwig2025} and multi-stage LWFA configurations~\cite{steinke2016, luo2018, haq2025}, can potentially yield higher energy gains, albeit at the cost of introducing additional degrees of freedom that substantially increase the complexity of the optimization problem. We defer their investigation to future work.

\section*{Results}

Under assumptions (i)–(iv), the LWFA problem can be fully described by only two dimensionless parameters: $ \mathcal{P}_0 / \mathcal{P}_{\mathrm{cr}} $ and $ \tau_0 \omega_{\mathrm{p}} $, where $ \tau_0 $ is the laser pulse duration (full width at half maximum of the intensity profile). The associated laser and plasma parameters can then be written in normalized form as~\cite{valenta2025, valenta2026}
\begin{gather}
    \label{eq:3}
    a_0 = 2 \left( \frac{\mathcal{P}_0}{\mathcal{P}_{\mathrm{cr}}} \right)^{1/3},\\
    \label{eq:4}
    \frac{w_0}{\lambda_0} = \frac{\sqrt{2}}{\uppi}
    \left( \frac{2 \sqrt{\uppi \ln{2}}}{\tau_0 \omega_{\mathrm{p}}} \right)^{1/3}
    \left( \frac{\mathcal{P}_0}{\mathcal{P}_{\mathrm{cr}}} \right)^{-1/6}
    \left( \frac{\mathcal{E}_0}{\overline{\mathcal{E}}} \right)^{1/3}, \\
    \label{eq:5}
    \frac{\tau_0}{T_0} = \sqrt{\frac{\ln{2}}{\uppi}} 
    \left( \frac{2 \sqrt{\uppi \ln{2}}}{\tau_0 \omega_{\mathrm{p}}} \right)^{-2/3}
    \left( \frac{\mathcal{P}_0}{\mathcal{P}_{\mathrm{cr}}} \right)^{-1/3}
    \left( \frac{\mathcal{E}_0}{\overline{\mathcal{E}}} \right)^{1/3},\\
    \label{eq:6}
    \frac{n_{\mathrm{e}}}{n_{\mathrm{cr}}} =
    \left( \frac{2 \sqrt{\uppi \ln{2}}}{\tau_0 \omega_{\mathrm{p}}} \right)^{-2/3}
    \left( \frac{\mathcal{P}_0}{\mathcal{P}_{\mathrm{cr}}} \right)^{2/3}
    \left( \frac{\mathcal{E}_0}{\overline{\mathcal{E}}} \right)^{-2/3}.
\end{gather}
Here, $ T_0 = \lambda_0 / c $ denotes the laser period and $ \overline{\mathcal{E}} = m_{\mathrm{e}} c^2 \lambda_0 / r_{\mathrm{e}} \approx 29~\mathrm{\upmu J} $ for $ \lambda_0 = 1~\mathrm{\upmu m} $ is a characteristic energy scale associated with electromagnetic waves interacting with electrons. In this work, we consider 
\begin{equation}\label{eq:7}
    0.1~\mathrm{J} \le \mathcal{E}_0 \le 1.6~\mathrm{J} \qquad \textnormal{and} \qquad 0.6~\mathrm{\upmu m} \le \lambda_0 \le 1.4~\mathrm{\upmu m},
\end{equation}
which covers the typical laser energies and wavelengths used in LWFA while also spanning a sufficiently large parameter region to allow extrapolation of the results.

\begin{figure}[t]
\centering
\includegraphics[width=0.65\textwidth]{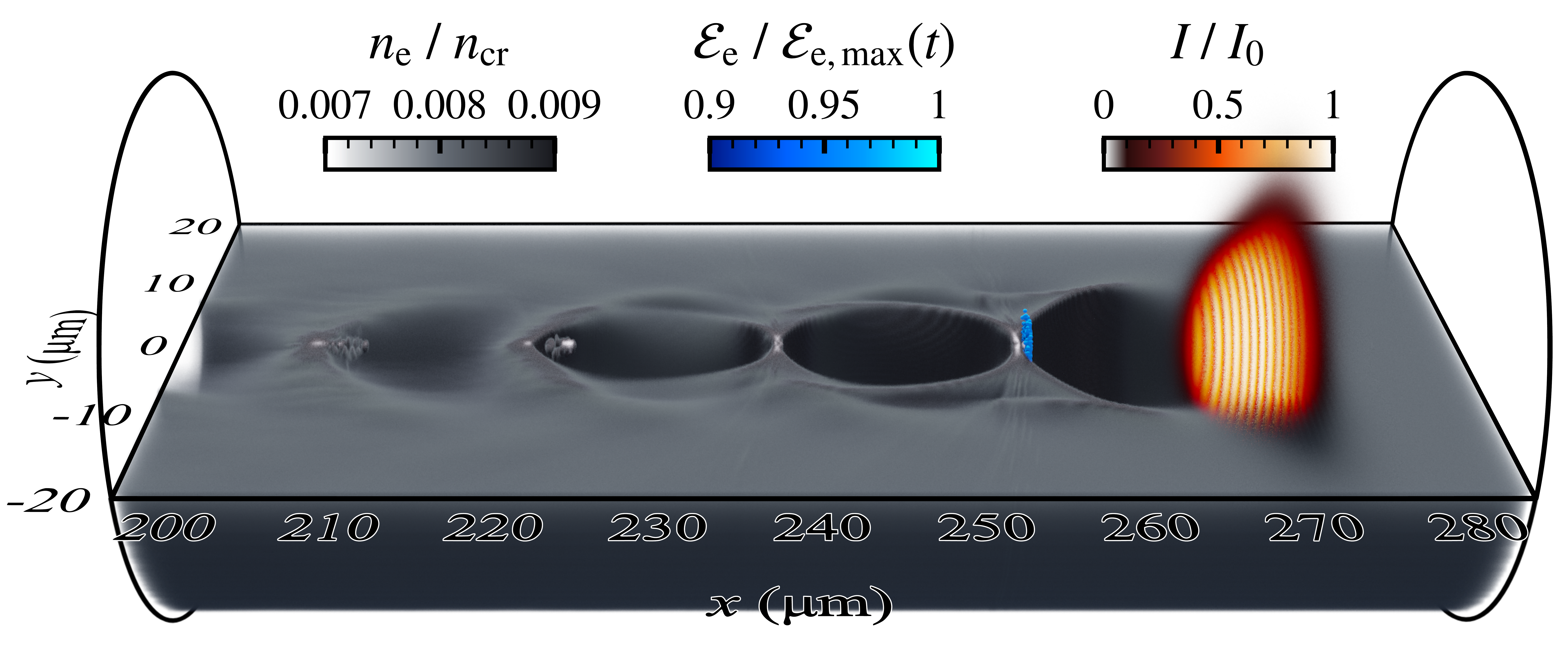}
\caption{\textbf{| Snapshot of a representative PIC simulation.} Spatial distributions of the laser intensity, $ I $, electron density, $ n_{\mathrm{e}} $, and energy of accelerated electrons, $ \mathcal{E}_{\mathrm{e}} $, at the time instant $ t \approx 300~T_0 $, obtained from the simulation with input parameters closest to the optimum identified by BO for a laser pulse with energy $ \mathcal{E}_0 = 0.4~\mathrm{J} $ and wavelength $ \lambda_0 = 1~\mathrm{\upmu m} $. Only test-electron macroparticles with energies within the upper $ 10\% $ of the instantaneous maximum electron energy, $ \mathcal{E}_{\mathrm{e,max}}(t) $, are shown. The plasma density is sliced along the $ x $–$ y $ plane to reveal the inner structure of the plasma wave and $ I_0 = 2 \mathcal{P}_0 / \uppi w_0^2 $.}
\label{fig:1}
\end{figure}

We systematically perform BO across a range of laser energies and wavelengths selected from the above-mentioned intervals. For each energy--wavelength pair, we construct surrogate models for the maximum electron energy, $ \mathcal{E}_{\mathrm{e,max}} $, and the corresponding shortest acceleration length, $ l_{\mathrm{acc}} $. Each surrogate model is based on the results of 128 PIC simulation trials. The parameters $ \mathcal{P}_0 / \mathcal{P}_{\mathrm{cr}} $ and $ \tau_0 \omega_{\mathrm{p}} $ are varied during optimization within the following fixed bounds: 
\begin{equation}\label{eq:8}
    2 \le \mathcal{P}_0 / \mathcal{P}_{\mathrm{cr}} \le 8 \qquad \textnormal{and} \qquad 1 \le \tau_0 \omega_{\mathrm{p}} \le 5.
\end{equation}
Based on our previous works~\cite{valenta2025, valenta2026}, the optimum is expected to lie within these intervals; hence the parameter limits are not dynamically adjusted during the optimization process. Further details of the PIC and BO setup are provided in the Methods.

\subsection*{Optimization with respect to laser energy}

\begin{figure}[t]
\centering
\includegraphics[width=0.9\textwidth]{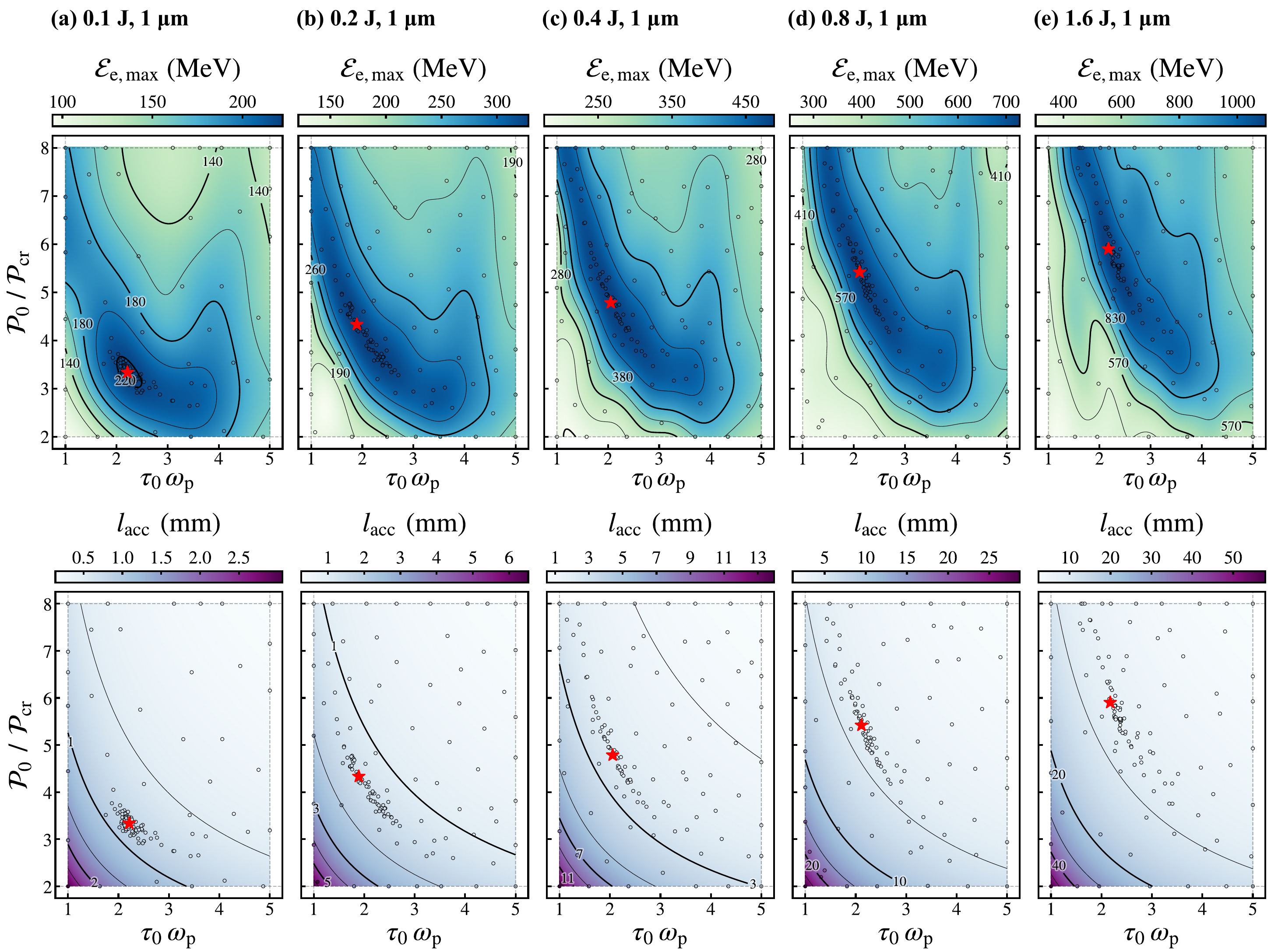}
\caption{\textbf{| Optimization with respect to laser energy.} Mean functions of the surrogate models for the maximum electron energy, $ \mathcal{E}_{\mathrm{e, max}} $, (upper plots) and the acceleration length, $ l_{\mathrm{acc}} $, (lower plots), each constructed from 128 PIC simulations at fixed laser wavelength, $ \lambda_0 = 1~\mathrm{\upmu m} $, and varying laser energy $ \mathcal{E}_0 $; \textbf{(a)} $ \mathcal{E}_0 = 0.1~\mathrm{J} $, \textbf{(b)} $ \mathcal{E}_0 = 0.2~\mathrm{J} $, \textbf{(c)} $ \mathcal{E}_0 = 0.4~\mathrm{J} $, \textbf{(d)} $ \mathcal{E}_0 = 0.8~\mathrm{J} $, and \textbf{(e)} $ \mathcal{E}_0 = 1.6~\mathrm{J} $. In each panel, the dots indicate the locations of the PIC simulation trials in parameter space $ ( \tau_0 \omega_{\mathrm{p}}, \mathcal{P}_0 / \mathcal{P}_{\mathrm{cr}} ) $, and the red star marks the point corresponding to the highest electron energy predicted by the surrogate model.}
\label{fig:2}
\end{figure}

First, we optimize the maximum electron energy with respect to laser energy at fixed laser wavelength, $ \lambda_0 = 1~\mathrm{\upmu m} $. The laser energy is varied as $ \mathcal{E}_0 = 0.1 $, $ 0.2 $, $ 0.4 $, $ 0.8 $, and $ 1.6~\mathrm{J} $. A representative optimized LWFA configuration obtained from a PIC simulation with $ \mathcal{E}_0 = 0.4~\mathrm{J} $ is shown in Fig.~\ref{fig:1}. The figure displays the laser intensity, plasma density, and accelerated electron beam at an intermediate stage of the acceleration process. The Gaussian process surrogate models for $ \mathcal{E}_{\mathrm{e,max}} $, together with the locations of the PIC simulation trials in the parameter space selected by the BO algorithm, are shown in the upper plots of Fig.~\ref{fig:2}. The surrogate models predict that the maximum values of $ \mathcal{E}_{\mathrm{e,max}} $, hereafter denoted by $ \mathcal{E}^{*} $, increase monotonically with laser energy, reaching $ \approx 222 $, $ 328 $, $ 488 $, $ 728 $, and $ 1098~\mathrm{MeV} $ for the above mentioned laser energies, respectively.

In addition to predicting the values of $ \mathcal{E}^{*} $, the surrogate models identify the parameter-space locations at which these optima occur, as indicated by the red stars in the upper plots of Fig.~\ref{fig:2}. The location of the optimum shifts systematically with laser energy. Importantly, however, it is not sharply localized; instead, near-maximum electron energies are obtained over a relatively broad region of parameter space. This suggests that the high-energy acceleration regime can be achieved for multiple combinations of initial laser and plasma parameters, consistent with the previous optimization study in the low-laser-energy limit~\cite{valenta2026}. Furthermore, except for the $0.1~\mathrm{J}$ case, this high-energy region retains a characteristic diagonal ridge in parameter space.

The lower plots of Fig.~\ref{fig:2} show the acceleration length corresponding to the maximum electron energy obtained in each PIC simulation trial, together with the associated surrogate models. The analysis of the acceleration length suggests a power-law dependence on the $ \mathcal{P}_0 / \mathcal{P}_{\mathrm{cr}} $ and $ \tau_0 \omega_{\mathrm{p}} $ parameters; the surrogate models for $ l_{\mathrm{acc}} $ are therefore constructed by fitting the simulation data with the model function $ f( \mathcal{P}_0 / \mathcal{P}_{\mathrm{cr}}, \tau_0 \omega_{\mathrm{p}}; \alpha, \beta, \gamma) = \alpha (\mathcal{P}_0 / \mathcal{P}_{\mathrm{cr}})^{\beta} (\tau_0 \omega_{\mathrm{p}})^{\gamma} $, where $ \alpha $, $ \beta $, and $ \gamma $ are adjustable parameters determined by the least-squares method.

The acceleration length corresponding to $ \mathcal{E}^{*} $, hereafter denoted by $ l^{*} $, is marked by the red stars in the lower plots of Fig.~\ref{fig:2}. This quantity is not an independent optimization objective but an analyzed parameter associated with the optimized electron energy $ \mathcal{E}^{*} $. Because $ \mathcal{E}^{*} $ is reached at this distance, $ l^{*} $ also represents the shortest acceleration length required to attain the optimized energy under the imposed assumptions. We do not attribute the acceleration length to individual limiting mechanisms (e.g., laser diffraction, depletion, or electron dephasing) as these processes are typically closely interconnected. As with $ \mathcal{E}^{*} $, the corresponding acceleration length $ l^{*} $ increases monotonically with laser energy; the surrogate models predict $ l^{*} \approx 0.8 $, $ 1.4 $, $ 2.3 $, $ 4.1 $, and $ 7.5~\mathrm{mm} $ for $ \mathcal{E}_0 = 0.1 $, $ 0.2 $, $ 0.4 $, $ 0.8 $, and $ 1.6~\mathrm{J} $, respectively.

\subsection*{Optimization with respect to laser wavelength}

\begin{figure}[t]
\centering
\includegraphics[width=0.9\textwidth]{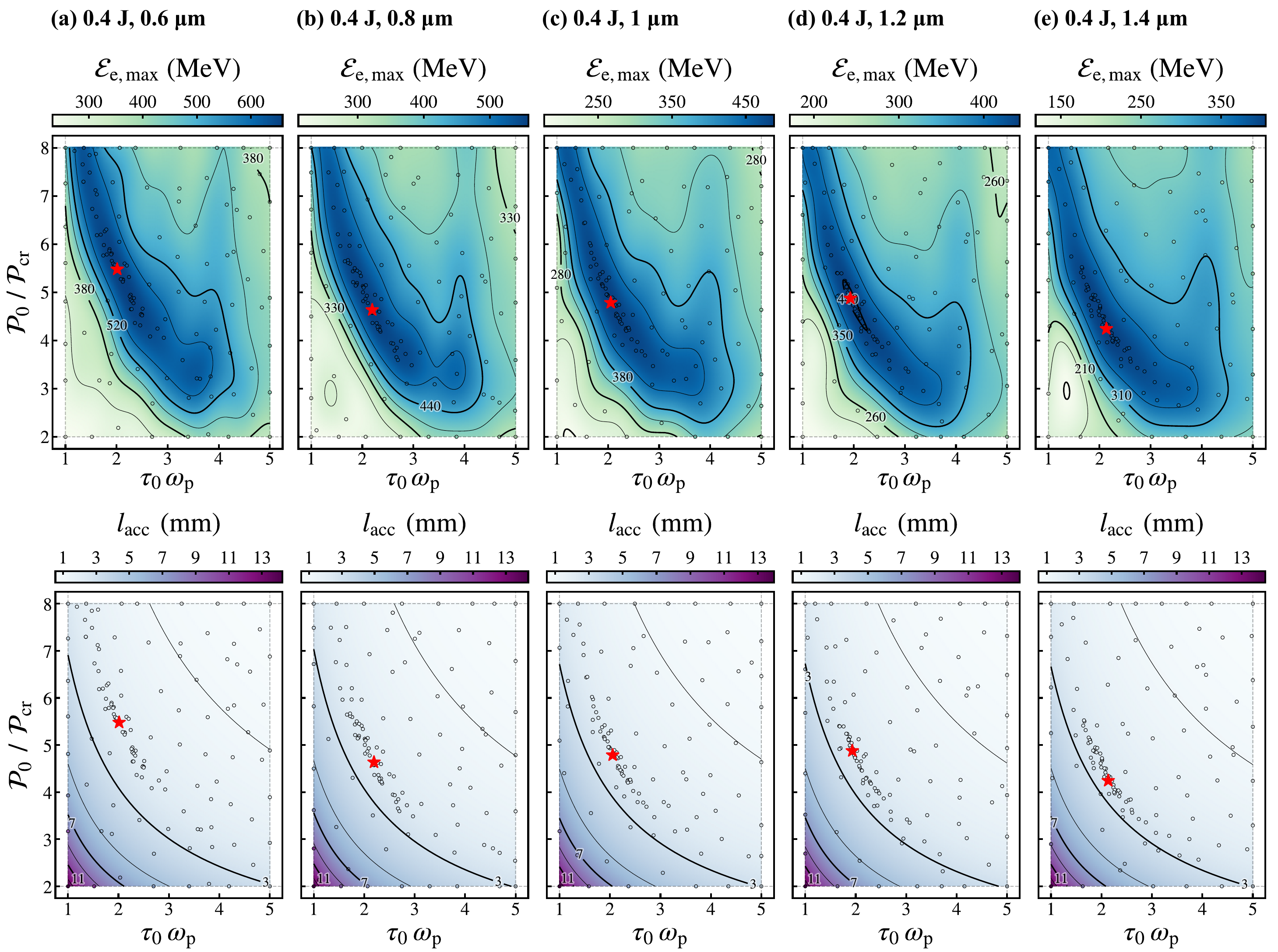}
\caption{\textbf{| Optimization with respect to laser wavelength.} Mean functions of the surrogate models for the maximum electron energy, $ \mathcal{E}_{\mathrm{e, max}} $, (upper plots) and the acceleration length, $ l_{\mathrm{acc}} $, (lower plots), each constructed from 128 PIC simulations at fixed laser energy, $ \mathcal{E}_0 = 0.4~\mathrm{J} $, and varying laser wavelength $ \lambda_0 $; \textbf{(a)} $ \lambda_0 = 0.6~\mathrm{\upmu m} $, \textbf{(b)} $ \lambda_0 = 0.8~\mathrm{\upmu m} $, \textbf{(c)} $ \lambda_0 = 1~\mathrm{\upmu m} $, \textbf{(d)} $ \lambda_0 = 1.2~\mathrm{\upmu m} $, and \textbf{(e)} $ \lambda_0 = 1.4~\mathrm{\upmu m} $. In each panel, the dots indicate the locations of the PIC simulation trials in parameter space $ ( \tau_0 \omega_{\mathrm{p}}, \mathcal{P}_0 / \mathcal{P}_{\mathrm{cr}} ) $, and the red star marks the point corresponding to the highest electron energy predicted by the surrogate model.}
\label{fig:3}
\end{figure}

Second, we optimize the maximum electron energy with respect to laser wavelength at fixed laser energy, $ \mathcal{E}_0 = 0.4~\mathrm{J} $. The laser wavelength is varied as $ \lambda_0 = 0.6 $, $ 0.8 $, $ 1.0 $, $ 1.2 $, and $ 1.4~\mathrm{\upmu m} $. The upper plots of Fig.~\ref{fig:3} display Gaussian process surrogate models for $ \mathcal{E}_{\mathrm{e,max}} $, together with the locations of the PIC simulation trials in the parameter space. The surrogate models predict that $ \mathcal{E}^{*} $ decreases monotonically with increasing laser wavelength, reaching $ \approx 659 $, $ 558 $, $ 488 $, $ 442 $, and $ 405~\mathrm{MeV} $ for the above mentioned wavelengths, respectively.

As in the optimization with respect to laser energy, the locations of the optima, indicated by the red stars in the upper plots of Fig.~\ref{fig:3}, shift with laser wavelength, although the trend is less pronounced. The characteristic shape of the high-energy region is preserved across all laser wavelengths considered. Together with the observed similarity across laser energies, this behavior supports the existence of scalable dynamics in the LWFA parameter space, motivating the development of energy-optimized scaling laws.

The acceleration length corresponding to the maximum electron energy obtained in each PIC simulation trial, together with the associated surrogate models, are shown in the lower plots of Fig.~\ref{fig:3}. As in the optimization with respect to laser energy, the surrogate models for $ l_{\mathrm{acc}} $ are obtained by fitting the simulation data with the same model function. Although $ \mathcal{E}^{*} $ decreases substantially with increasing laser wavelength, the corresponding acceleration length varies only weakly. The surrogate models predict $ l^{*} \approx 2.1 $, $ 2.3 $, $ 2.3 $, $ 2.4 $, and $ 2.6~\mathrm{mm} $ for $ \lambda_0 = 0.6 $, $ 0.8 $, $ 1.0 $, $ 1.2 $, and $ 1.4~\mathrm{\upmu m} $, respectively.

\subsection*{Energy-optimized scaling laws}

\begin{figure}[t]
\centering
\includegraphics[width=0.85\textwidth]{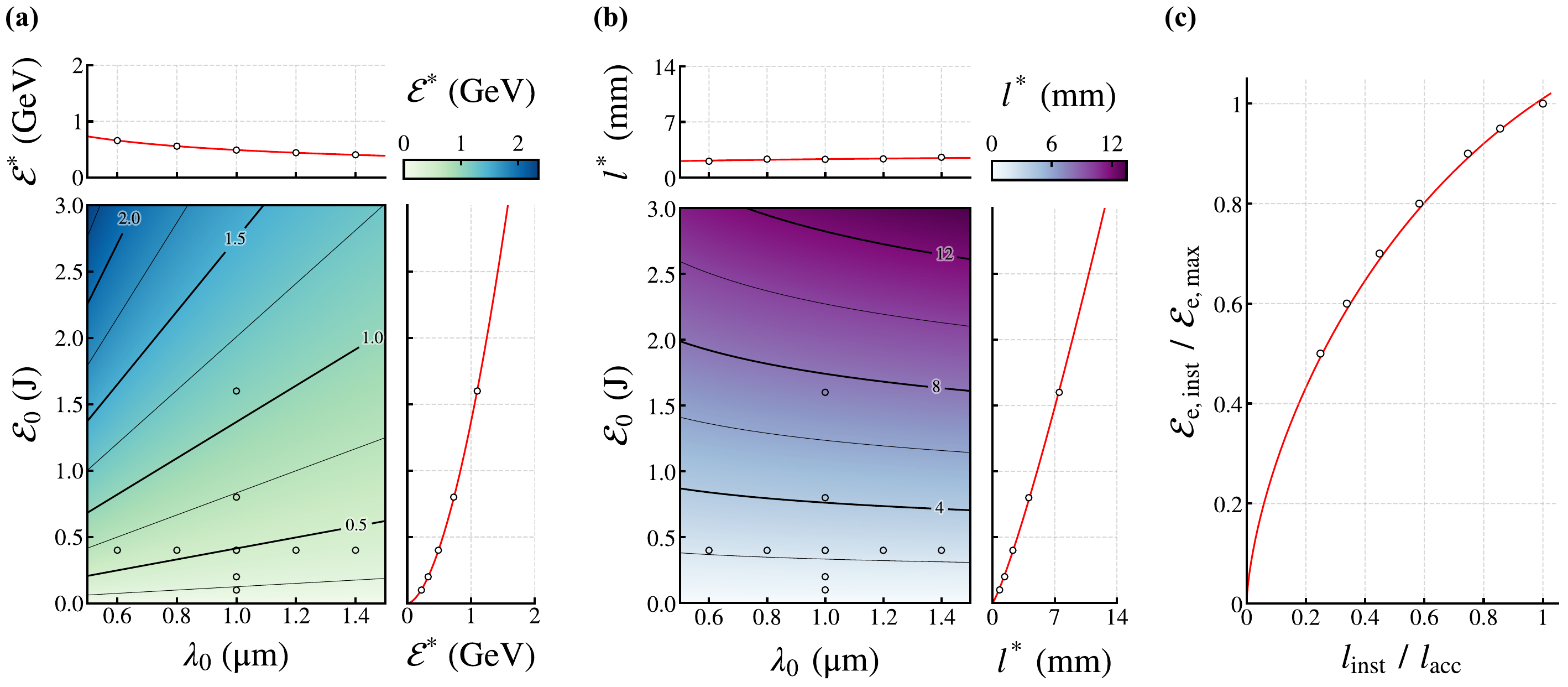}
\caption{\textbf{| Energy-optimized scaling laws.} \textbf{(a)} Maximum electron energy, $ \mathcal{E}^{*} $, and \textbf{(b)} the corresponding shortest acceleration length, $ l^{*} $, in self-guided LWFA as functions of laser energy, $ \mathcal{E}_0 $, and wavelength, $ \lambda_0 $ given by Eqs.~\eqref{eq:9}. In panels \textbf{(a)} and \textbf{(b)}, the upper plots show line-outs at fixed laser energy, $\mathcal{E}_0 = 0.4~\mathrm{J}$, whereas the right plots show line-outs at fixed laser wavelength, $\lambda_0 = 1~\mathrm{\upmu m}$; in both panels, the dots denote the optimal values predicted by the surrogate models, while the solid red lines show the analytical approximations given by Eqs.~\eqref{eq:9}. \textbf{(c)} Mean instantaneous acceleration length, $ l_{\mathrm{inst}} $, relative to the total acceleration length, $ l_{\mathrm{acc}} $, at the corresponding fractions of the maximum electron energy, $ \mathcal{E}_{\mathrm{e, inst}} / \mathcal{E}_{\mathrm{e, max}} $, averaged over all PIC simulation trials (black dots), together with the approximation given by Eq.~\eqref{eq:10} (solid red line).}
\label{fig:4}
\end{figure}

By extracting the values of $ \mathcal{E}^{*} $ and $ l^{*} $ from the surrogate models shown in Figs.~\ref{fig:2} and \ref{fig:3} and fitting them using the least-squares method with the model function $ f(\mathcal{E}_0 / \overline{\mathcal{E}}; \alpha, \beta) = \alpha (\mathcal{E}_0 / \overline{\mathcal{E}})^{\beta} $, where $ \alpha $ and $ \beta $ are adjustable parameters, we obtain the following energy-optimized scaling laws for self-guided LWFA:
\begin{equation}\label{eq:9}
    \frac{\mathcal{E}^{*}}{m_{\mathrm{e}} c^2} \approx 3.81 \left( \frac{\mathcal{E}_0}{\overline{\mathcal{E}}} \right)^{0.58} \qquad \textnormal{and} \qquad \frac{l^{*}}{\lambda_0} \approx 0.78 \left( \frac{\mathcal{E}_0}{\overline{\mathcal{E}}} \right)^{0.84}.
\end{equation}
The optimal values extracted from the surrogate models, together with the approximations given by Eqs.~\eqref{eq:9}, are both shown in panels (a) and (b) of Fig.~\ref{fig:4}. The obtained scaling laws imply that doubling the laser energy increases the maximum electron energy by a factor of $ \approx 1.5 $, while increasing the acceleration length by a factor of $ \approx 1.8 $. An equivalent increase in electron energy can be achieved by halving the laser wavelength, but in this case the required acceleration length decreases by a factor of $ \approx 0.9 $. Applying the scaling laws to, e.g., a $ 1~\mathrm{J} $, $ 1~\mathrm{\upmu m} $ laser driver yields electrons with maximum energy of $ \approx 833~\mathrm{MeV} $ accelerated over a length of $ \approx 5.1~\mathrm{mm} $.

Furthermore, for each PIC simulation trial, we record the acceleration length at which the electrons reach $50$, $60$, $70$, $80$, $90$, and $95\%$ of their maximum energy. This allows us to establish the relationship between the instantaneous electron energy, $ \mathcal{E}_{\mathrm{e,inst}} $, and the corresponding instantaneous acceleration length, $ l_{\mathrm{inst}} $. Motivated by the phase-space trajectories of trapped electrons~\cite{esirkepov2006}, we find that this relationship is well described by a least-squares fit to a model function representing a cycloid segment,
\begin{equation}\label{eq:10}
    \frac{l_{\mathrm{inst}}}{l_{\mathrm{acc}}} = \alpha \arccos \left(1 - \alpha^{-1} \frac{\mathcal{E}_{\mathrm{e, inst}}}{\mathcal{E}_{\mathrm{e, max}}} \right) - \sqrt{ \frac{\mathcal{E}_{\mathrm{e, inst}}}{\mathcal{E}_{\mathrm{e, max}}} \left(2 \alpha  - \frac{\mathcal{E}_{\mathrm{e, inst}}}{\mathcal{E}_{\mathrm{e, max}}} \right)},
\end{equation}
where $ \alpha \approx 0.58 $. Figure~\ref{fig:4}(c) shows the mean values of $ l_{\mathrm{inst}} $ relative to $ l_{\mathrm{acc}} $ at the corresponding fractions of the maximum electron energy, averaged over all PIC simulation trials, together with the approximation given by Eq.~\eqref{eq:10}. This result is particularly useful in practice: electrons on average reach half of their maximum energy after only $ 26 \% $ of the total acceleration length, and $ 75 \% $ after only $ 53 \% $ of $ l_{\mathrm{acc}} $. Consequently, doubling the laser energy reduces the acceleration length required to reach the original maximum electron energy by a factor of $ \approx 1.3 $, whereas halving the laser wavelength reduces the required acceleration length by a factor of $ \approx 2.6 $.

\subsection*{Optimal input parameters}

To identify the laser and plasma parameters underlying the scaling laws given by Eqs.~\eqref{eq:9}, we track the location of the optimal point (i.e., the one that yields the maximum electron energy) in the parameter space. Figures~\ref{fig:2} and \ref{fig:3} show that the optimal value of the parameter $ \mathcal{P}_0 / \mathcal{P}_{\mathrm{cr}} $, hereafter denoted by $ \mathcal{P}^{*} $, drifts toward higher values as the laser energy increases and the laser wavelength decreases. A least-squares fit using the model function $ f(\mathcal{E}_0 / \overline{\mathcal{E}}; \alpha, \beta) = \alpha (\mathcal{E}_0 / \overline{\mathcal{E}})^{\beta} $, where $ \alpha $ and $ \beta $ are adjustable parameters, yields the following dependence,
\begin{equation}\label{eq:11}
    \mathcal{P}^{*} \approx 0.79 \left(\frac{\mathcal{E}_0}{\overline{\mathcal{E}}} \right)^{0.19}.
\end{equation}

On the other hand, the optimal value of the parameter $ \tau_0 \omega_{\mathrm{p}} $, hereafter denoted by $ \tau^{*} $, varies only weakly across the range of laser energies and wavelengths considered. A closer examination reveals a slight increase of $ \tau^{*} $ with increasing laser energy. At lower laser energies, however, this trend appears to reverse, with $ \tau^{*} \approx 2.2 $ at $100~\mathrm{mJ}$ (and $ \tau^{*} \approx 3 $ at $10~\mathrm{mJ}$~\cite{valenta2026}), coinciding with the qualitative change in the surrogate model observed in the upper plot of Fig.~\ref{fig:2}(a). Excluding the $0.1~\mathrm{J}$ case, a least-squares fit to the optimal values obtained for the remaining laser energy--wavelength pairs using the model function $ f(\mathcal{E}_0 / \overline{\mathcal{E}}; \alpha, \beta) = \alpha (\mathcal{E}_0 / \overline{\mathcal{E}})^{\beta} $, where $ \alpha $ and $ \beta $ are adjustable parameters, yields
\begin{equation}
\label{eq:12}
    \tau^{*} \approx 1.29 \left(\frac{\mathcal{E}_0}{\overline{\mathcal{E}}} \right)^{0.05}.
\end{equation}

The complete set of optimal input laser and plasma parameters that enable the scaling laws given by Eqs.~\eqref{eq:9} can be obtained by replacing $\tau_0\omega_{\mathrm{p}}$ and $\mathcal{P}_0/\mathcal{P}_{\mathrm{cr}}$ in Eqs.~\eqref{eq:3}--\eqref{eq:6} with the optimal values given by Eqs.~\eqref{eq:11} and \eqref{eq:12}, respectively,
\begin{gather}
    \label{eq:13}
    a_{0, \mathrm{opt}} \approx 1.85 \left( \frac{\mathcal{E}_0}{\overline{\mathcal{E}}} \right)^{0.06},\\
    \label{eq:14}
    \frac{w_{0, \mathrm{opt}}}{\lambda_0} \approx 0.62
    \left( \frac{\mathcal{E}_0}{\overline{\mathcal{E}}} \right)^{0.29}, \\
    \label{eq:15}
    \frac{\tau_{0, \mathrm{opt}}}{T_0} \approx 0.29
    \left( \frac{\mathcal{E}_0}{\overline{\mathcal{E}}} \right)^{0.3},\\
    \label{eq:16}
    \frac{n_{\mathrm{e, opt}}}{n_{\mathrm{cr}}} \approx 0.49
    \left( \frac{\mathcal{E}_0}{\overline{\mathcal{E}}} \right)^{-0.51}.
\end{gather}
According to Eqs.~\eqref{eq:13}--\eqref{eq:16}, the optimal input parameters for, e.g., a $ 1~\mathrm{J} $, $ 1~\mathrm{\upmu m} $ laser driver are $ a_{0, \mathrm{opt}} \approx 3.5 $, $ w_{0, \mathrm{opt}} \approx 12.8~\mathrm{\upmu m} $, $ \tau_{0, \mathrm{opt}} \approx 22.2~\mathrm{fs} $, and $ n_{\mathrm{e, opt}} \approx 2.7 \times 10^{18}~\mathrm{cm^{-3}} $. An overview of the optimized parameters for all investigated laser energy--wavelength pairs is provided in Tab.~\ref{tab:1}.

\begin{table}[ht]
\centering
\begin{tabular}{cccccccccc}
\hline
\hline
$ \mathcal{E}_0~(\mathrm{J}) $ & $ \lambda_0~(\mathrm{\upmu m}) $ & $ \mathcal{E}^{*}~(\mathrm{MeV}) $ & $ l^{*}~(\mathrm{mm}) $ & $ \mathcal{P}^{*} $ & $ \tau^{*} $ & $ a_{0, \mathrm{opt}} $ & $ w_{0, \mathrm{opt}}~(\mathrm{\upmu m}) $ & $ \tau_{0, \mathrm{opt}}~(\mathrm{fs}) $ & $ n_{\mathrm{e, opt}}~(10^{18}~\mathrm{cm^{-3}}) $ \\
\hline

$ 0.1 $ & $ 1.0 $ & $ 222 $ & $ 0.8 $ & $ 3.33 $ & $ 2.23 $ & $ 3.07 $ & $ 6.33 $ & $ 11.56 $ & $ 8.67 $ \\

$ 0.2 $ & $ 1.0 $ & $ 328 $ & $ 1.4 $ & $ 4.32 $ & $ 1.90 $ & $ 3.21 $ & $ 7.71 $ & $ 14.27 $ & $ 6.09 $ \\

$ 0.4 $ & $ 1.0 $ & $ 488 $ & $ 2.3 $ & $ 4.77 $ & $ 2.07 $ & $ 3.35 $ & $ 9.41 $ & $ 17.61 $ & $ 4.28 $ \\

$ 0.8 $ & $ 1.0 $ & $ 728 $ & $ 4.1 $ & $ 5.41 $ & $ 2.13 $ & $ 3.50 $ & $ 11.47 $ & $ 21.73 $ & $ 3.00 $ \\

$ 1.6 $ & $ 1.0 $ & $ 1098 $ & $ 7.5 $ & $ 5.89 $ & $ 2.19 $ & $ 3.65 $ & $ 13.98 $ & $ 26.82 $ & $ 2.11 $ \\

$ 0.4 $ & $ 0.6 $ & $ 659 $ & $ 2.1 $ & $ 5.47 $ & $ 2.03 $ & $ 3.46 $ & $ 6.53 $ & $ 12.34 $ & $ 9.16 $ \\

$ 0.4 $ & $ 0.8 $ & $ 558 $ & $ 2.3 $ & $ 4.62 $ & $ 2.21 $ & $ 3.40 $ & $ 8.02 $ & $ 15.07 $ & $ 5.96 $ \\

$ 0.4 $ & $ 1.2 $ & $ 442 $ & $ 2.4 $ & $ 4.86 $ & $ 1.95 $ & $ 3.31 $ & $ 10.71 $ & $ 19.99 $ & $ 3.26 $ \\

$ 0.4 $ & $ 1.4 $ & $ 405 $ & $ 2.6 $ & $ 4.23 $ & $ 2.15 $ & $ 3.28 $ & $ 11.96 $ & $ 22.26 $ & $ 2.59 $ \\
\hline
\hline
\end{tabular}
\caption{\textbf{| Overview of the optimized parameters.} Maximum electron energy, $\mathcal{E}^{*}$; the corresponding acceleration length, $l^{*}$; the values of $\mathcal{P}^{*}$ and $\tau^{*}$ parameters; and the corresponding optimized laser and plasma input parameters, namely the normalized laser amplitude, $a_{0,\mathrm{opt}}$, beam waist, $w_{0,\mathrm{opt}}$, pulse duration, $\tau_{0,\mathrm{opt}}$, and plasma electron density, $n_{\mathrm{e,opt}}$, for each laser energy, $\mathcal{E}_0$, and wavelength, $\lambda_0$, considered in this work.}
\label{tab:1}
\end{table}

\section*{Discussion}

In Ref.~\citenum{lu2007}, the scaling laws for the maximum electron energy and the corresponding acceleration length in self-guided LWFA were estimated (neglecting numerical prefactors) as
\begin{equation}\label{eq:17}
\frac{\mathcal{E}^{*}}{m_{\mathrm{e}} c^2} \sim a_0 \left( \frac{n_{\mathrm{e}}}{n_{\mathrm{cr}}} \right)^{-1} \qquad \textnormal{and} \qquad \frac{l^{*}}{\lambda_0} \sim \sqrt{a_0} \left( \frac{n_{\mathrm{e}}}{n_{\mathrm{cr}}} \right)^{-3 / 2}.
\end{equation}
Substituting the expressions for $a_0$ and $n_{\mathrm{e}}/n_{\mathrm{cr}}$ from Eqs.~\eqref{eq:13} and \eqref{eq:16}, respectively, we recover relatively precisely the scaling laws of Eqs.~\eqref{eq:9} (up to the numerical prefactors), $\mathcal{E}^{*}/m_{\mathrm{e}}c^2 \sim (\mathcal{E}_0/\overline{\mathcal{E}})^{0.57}$ and $l^{*}/\lambda_0 \sim (\mathcal{E}_0/\overline{\mathcal{E}})^{0.8}$. The scaling exponents in Eqs.~\eqref{eq:9} thus follow directly from the analytical model of Ref.~\citenum{lu2007}. In contrast, the numerical prefactors, as determined from the PIC simulations, naturally account for the full nonlinear evolution of the wakefield and laser pulse, thereby incorporating effects that are only approximately represented in analytical models (e.g., the pronounced accelerating-field spike at the rear of the plasma cavity, which provides an additional energy boost during the early stages of acceleration, and the subsequent depletion of the laser pulse, which ultimately terminates the acceleration once self-guiding can no longer be sustained).

The optimal values of the laser and plasma parameters underlying the scaling laws given by Eqs.~\eqref{eq:13}--\eqref{eq:16} are also physically consistent with previous theoretical predictions. The phenomenological theory of self-guided LWFA predicts that the laser amplitude must increase as the plasma density decreases to sustain efficient self-guided LWFA~\cite{lu2007, davidson2019}. This is because the relativistic increase of the refractive index at the front of the laser pulse is compensated by the density compression produced by the ponderomotive force. Stable self-guiding becomes possible only when the laser amplitude is sufficiently high that the leading edge of the pulse locally pump-depletes while exciting the wakefield, allowing the undepleted portion of the pulse, located behind the density compression, to remain guided. The authors of Ref.~\citenum{lu2007} estimate the optimal value as $ a_0 \sim (n_\mathrm{cr} / n_{\mathrm{e}})^{1/5} $, which can be equivalently expressed as $ \mathcal{P}^{*} \sim (\mathcal{E}_0 / \overline{\mathcal{E}})^{2/7} $. This exponent is in relatively good agreement with the value obtained from the least-squares fit in Eq.~\eqref{eq:11}.

It is well established that the wakefield amplitude is maximized when the laser pulse duration matches the resonant duration for plasma-wave excitation~\cite{leemans1996}. As the plasma wave wavelength increases not only with decreasing plasma density but also with increasing laser amplitude, the optimal value of the normalized pulse duration, $ \tau^{*} $, is also expected to increase with $ a_0 $. Analytical estimates~\cite{bulanov2016} predict $ \tau^{*} \sim a_0 $, whereas the least-squares fit in Eq.~\eqref{eq:12} yields $ \tau^{*} \sim a_0^{0.8} $, again in relatively good agreement with the analytical prediction.

As discussed above, a significant fraction of the leading edge of the laser pulse undergoes local pump depletion while driving the wakefield before diffraction becomes dominant. As the leading edge erodes, the effective pulse length is progressively reduced, implying a lower limit to the normalized pulse duration below which the leading edge constitutes an increasingly large fraction of the pulse. Once this limit is reached, pump depletion is no longer sufficient to compensate diffraction, preventing stable self-guiding. We attribute the shift of the optimum toward pulse durations longer than predicted by Eq.~\eqref{eq:12} observed at low laser energies to this effect [see the upper plot of Fig.~\ref{fig:2}(a) and Ref.~\citenum{valenta2026}]. For example, for a $0.1~\mathrm{J}$ driver pulse, the region defined by $ 5 \le \mathcal{P}_0 / \mathcal{P}_{\mathrm{cr}} \le 8 $ and $ 1 \le \tau_0 \omega_{\mathrm{p}} \le 2 $, which otherwise supports high-electron-energy LWFA at higher driver energies, already corresponds to very short pulse durations ($\approx6$--$11~\mathrm{fs}$). Furthermore, under these conditions, carrier-envelope-phase effects may become significant~\cite{nerush2009,valenta2020,huijts2021}, potentially contributing to the observed deviation.

To assess the accuracy of the scaling laws given by Eqs.~\eqref{eq:9}, we perform additional higher-fidelity quasi-3D laboratory-frame simulations for each identified optimum. Compared with the original Lorentz-boosted-frame simulations, these simulations predict maximum electron energies higher by a factor of $ \approx 1.08 $--$ 1.16 $ (with $ \approx 1.16 $ at low $ \mathcal{E}_0 $ and $ \approx 1.08 $ at high $ \mathcal{E}_0 $) and corresponding acceleration lengths shorter by a factor of $ \approx 0.91 $--$ 0.97 $ (with $ \approx 0.97 $ at low $ \mathcal{E}_0 $ and $ \approx 0.91 $ at high $ \mathcal{E}_0 $). These differences suggest that the scaling exponent in the first of Eqs.~\eqref{eq:9} may decrease by $ \approx \ln{(1.08 / 1.16)} / \ln{(1.6 / 0.1)} \approx -0.03 $ while the corresponding prefactor may increase by $ \approx 1.16 (0.1 / 29 \times 10^{-6})^{0.03} \approx 1.48 $. Similarly, the scaling exponent in the second of Eqs.~\eqref{eq:9} may decrease by $ \approx \ln{(0.91 / 0.97)} / \ln{(1.6 / 0.1)} \approx -0.02 $, whereas the corresponding prefactor may increase by $ \approx 0.97 (0.1 / 29 \times 10^{-6})^{0.02} \approx 1.14 $.

Finally, we emphasize that the scaling laws of Eqs.~\eqref{eq:9} and the corresponding optimal input parameters given by Eqs.~\eqref{eq:13}--\eqref{eq:16} were derived for the laser energies and wavelengths specified by Eq.~\eqref{eq:7}, covering the range typically employed in present-day LWFA experiments. Their extrapolation beyond this range should therefore be treated with caution, as illustrated by the deviations observed in the low-energy limit discussed above.

In summary, we introduce a framework that combines BO with a large-scale PIC simulation study to derive generalized, energy-based scaling laws for self-guided LWFA driven by lasers of arbitrary energy and wavelength. These scaling laws are explicitly optimized to provide the maximum electron energy over the shortest acceleration length and are accompanied by the complete set of input parameters required for their realization. Our results provide practical guidance for the design of state-of-the-art LWFA experiments operating at their fundamental performance limits and demonstrate that increasing the laser energy, rather than the peak power, or alternatively decreasing the laser wavelength, is a key route toward extending LWFA to collider-relevant parameter regimes.

\section*{Methods}

\subsection*{Particle-in-cell simulations}

The driving laser pulse in the PIC simulations has the energy of $ \mathcal{E}_0 $ and a wavelength of $ \lambda_0 $. It has Gaussian spatial and temporal profiles, characterized by the FWHM duration $ \tau_0 $, normalized amplitude $ a_0 $, and beam waist $ w_0 $. It propagates along the $ x $ axis, is linearly polarized along the $ y $ axis, and interacts with a fully ionized plasma slab of uniform electron density $ n_{\mathrm{e}} $. A 10 $\mathrm{\upmu m}$-long density ramp is applied at the plasma entrance to suppress plasma-wave breaking and spurious electron injection at the sharp plasma–vacuum interface. The laser is focused at the end of the density ramp. The parameters $ \tau_0 $, $ a_0 $, $ w_0 $, and $ n_{\mathrm{e}} $ are updated iteratively according to Eqs.~\eqref{eq:3}--\eqref{eq:6} during BO.

To isolate the acceleration dynamics from the specifics of individual injection mechanisms, we employ an externally injected beam of test electrons. The test electrons evolve self-consistently under the electromagnetic fields but do not contribute current to the plasma, thereby acting as passive tracers. This approximation is valid well below the beam-loading threshold~\cite{wilks1987,katsouleas1987, tzoufras2009}, where the accelerated charge has a negligible effect on the wakefield. The initial velocity of the test electrons is set to approximately match the phase velocity of the plasma wave and thus to ensure efficient electron trapping, $ v_0 = c \sqrt{1 - \omega_{\mathrm{p}}^2 / \omega_0^2} $. The beam is initialized as a cylinder extending over the entire simulation domain with radius $ 2 w_0 $, allowing a continuous range of injection phases to be sampled within a single simulation. Using passive test electrons allows the acceleration process to be optimized independently of the injection mechanism, making the resulting scaling laws broadly applicable.

The simulations are performed with \textsc{Osiris}~\cite{fonseca2002} using the quasi-3D geometry based on an azimuthal Fourier decomposition of the electromagnetic fields~\cite{lifschitz2009,davidson2015}. Two azimuthal modes are retained, describing the axisymmetric plasma response and the non-axisymmetric linearly polarized laser field. A moving window propagating at the speed of light is employed, with dimensions of $ 4 (\mathcal{E}_0 / \overline{\mathcal{E}})^{1/3} \lambda_0 $ and $ 2.5 (\mathcal{E}_0 / \overline{\mathcal{E}})^{1/3} \lambda_0 $ and cell sizes of $ k_0 \Delta x = 2 \uppi / 10 $ and $ k_{\mathrm{p}} \Delta r = 2 \uppi / 100 $ in the longitudinal and radial directions, respectively, where $ k_0 = \omega_0 / c $ denotes the laser wavenumber. Each simulation is evolved for $ (n_{\mathrm{e}} / n_{\mathrm{cr}})^{-3/2} T_0 $, sufficient to capture the maximum electron energy in all cases considered. To reduce computational cost, the bulk of the simulations are performed in an optimally chosen Lorentz-boosted frame~\cite{vay2007,fonseca2008,yu2016}. Both the quasi-3D formulation and Lorentz-boosted-frame technique are well established for LWFA simulations~\cite{martins2010, yu2014, yu2016, ludwig2025, massimo2025}.

The background plasma is assumed to be cold and collisionless and is represented by electron and ion macroparticles with cubic shape functions. Each cell initially contains 32 background-electron, 32 test-electron, and 8 ion macroparticles. Macroparticle trajectories are advanced with the Boris pusher~\cite{boris1971}, and the electromagnetic fields are evolved using a finite-difference time-domain Maxwell solver optimized for Lorentz-boosted-frame simulations~\cite{li2017}.

In addition, higher-fidelity quasi-3D laboratory-frame simulations are performed for each identified optimum with cell sizes of $ k_0 \Delta x = 2 \uppi / 30 $ and $ k_{\mathrm{p}} \Delta r = 2 \uppi / 300 $, while keeping all other simulation parameters unchanged.

\subsection*{Bayesian optimization}

To efficiently explore the parameter space while minimizing the number of PIC simulations, we couple the simulations with BO. The optimization is performed using \textsc{Optimas}, a framework for scalable optimization on high-performance computing systems~\cite{hudson2022,ferranpousa2023}. BO has previously been applied successfully to a variety of optimization problems in LWFA~\cite{shalloo2020, jalas2021, kirchen2021, jalas2023, irshad2023, ferranpousa2023, irshad2024, nunes2025, valenta2025, valenta2026, djordjevic2026, maslarova2026}.

BO iteratively constructs a probabilistic surrogate model of the objective function and uses an acquisition function to select subsequent simulation trials. The surrogate model provides both a prediction of the objective function and an associated uncertainty, while the acquisition function balances exploration of uncertain regions with exploitation of promising candidates. As new simulation results become available, the surrogate model is updated, enabling efficient identification of the optimum with substantially fewer evaluations than exhaustive or random searches.

We employ a Gaussian process~\cite{rasmussen2005} with a radial basis function kernel~\cite{powell1977} as the surrogate model, motivated by the expected smooth dependence of the output on the input parameters. The acquisition function is the Monte Carlo batched upper confidence bound ($q$-UCB)~\cite{auer2002,wilson2017}, with the hyperparameter $ \beta_{q\text{-}\mathrm{UCB}} = 10 $, which provides appropriate balance between exploration and exploitation within the available simulation budget. Batched acquisition enables multiple trials to be evaluated concurrently.

The single optimization objective is the maximum electron energy attained by electrons in the first plasma-wave period behind the driver pulse. The maximum electron energy is defined as the energy reached by the test-electron macroparticles in the top $ 1\% $ of the energy distribution. Additionally, the acceleration length, defined as the propagation distance in plasma at which the maximum electron energy is attained, is recorded for post-analysis but is not included in the optimization objective.

The optimization follows a distributed asynchronous workflow. A single manager process generates and collects trials, while eight worker processes perform independent PIC simulations in parallel. The asynchronous execution prevents idle resources caused by variations in simulation runtime. Communication is implemented using the Message Passing Interface (MPI)~\cite{clarke1994}. The manager runs on a dedicated compute node, whereas simulations use two to four nodes, each with 32 MPI processes and 4 threads per process.

The optimization is initialized with eight Sobol samples~\cite{sobol1967} and terminated after a total budget of 128 trials. By this stage, the sampled points form a dense cluster around the optimum, and the objective values obtained in the final iterations exhibit negligible variation, indicating convergence.

\section*{Data availability}

All data supporting the findings of this study are available from the corresponding author upon request.

\section*{Acknowledgements}
This work was supported by the Ministry of Education, Youth and Sports of the Czech Republic through the e-INFRA CZ (ID:90254). K.G.M. discloses support for the research of this work from the DOE Office of Fusion Energy Sciences under Award Number DE-SC0021057 and the DOE National Nuclear Security Administration (NNSA) through the University of Rochester’s ‘National Inertial Confinement Fusion Program’ under Award Number DE-NA0004144.

\section*{Competing interests}

The authors declare no competing interests.


\begin{thebibliography}{100}
\urlstyle{rm}
\expandafter\ifx\csname url\endcsname\relax
  \def\url#1{\texttt{#1}}\fi
\expandafter\ifx\csname urlprefix\endcsname\relax\def\urlprefix{URL }\fi
\expandafter\ifx\csname doiprefix\endcsname\relax\def\doiprefix{DOI: }\fi
\providecommand{\bibinfo}[2]{#2}
\providecommand{\eprint}[2][]{\url{#2}}

\bibitem{tajima1979}
\bibinfo{author}{Tajima, T.} \& \bibinfo{author}{Dawson, J.~M.}
\newblock \bibinfo{journal}{\bibinfo{title}{Laser electron accelerator}}.
\newblock {\emph{\JournalTitle{Physical Review Letters}}} \textbf{\bibinfo{volume}{43}}, \bibinfo{pages}{267--270}, \doiprefix\url{10.1103/PhysRevLett.43.267} (\bibinfo{year}{1979}).

\bibitem{malka2008}
\bibinfo{author}{Malka, V.} \emph{et~al.}
\newblock \bibinfo{journal}{\bibinfo{title}{Principles and applications of compact laser--plasma accelerators}}.
\newblock {\emph{\JournalTitle{Nature Physics}}} \textbf{\bibinfo{volume}{4}}, \bibinfo{pages}{447--453}, \doiprefix\url{10.1038/nphys966} (\bibinfo{year}{2008}).

\bibitem{esarey2009}
\bibinfo{author}{Esarey, E.}, \bibinfo{author}{Schroeder, C.~B.} \& \bibinfo{author}{Leemans, W.~P.}
\newblock \bibinfo{journal}{\bibinfo{title}{Physics of laser-driven plasma-based electron accelerators}}.
\newblock {\emph{\JournalTitle{Reviews of Modern Physics}}} \textbf{\bibinfo{volume}{81}}, \bibinfo{pages}{1229--1285}, \doiprefix\url{10.1103/RevModPhys.81.1229} (\bibinfo{year}{2009}).

\bibitem{bulanov2016}
\bibinfo{author}{Bulanov, S.~V.} \emph{et~al.}
\newblock \bibinfo{journal}{\bibinfo{title}{On some theoretical problems of laser wake-field accelerators}}.
\newblock {\emph{\JournalTitle{Journal of Plasma Physics}}} \textbf{\bibinfo{volume}{82}}, \bibinfo{pages}{905820308}, \doiprefix\url{10.1017/S0022377816000623} (\bibinfo{year}{2016}).

\bibitem{kato2026}
\bibinfo{author}{Kato, Y.}, \bibinfo{author}{Mima, K.} \& \bibinfo{author}{Bulanov, S.}
\newblock \emph{\bibinfo{title}{High {{Power Laser}} and {{Plasma Science}}}}, vol. \bibinfo{volume}{130} of \emph{\bibinfo{series}{Springer {{Series}} on {{Atomic}}, {{Optical}}, and {{Plasma Physics}}}} (\bibinfo{publisher}{Springer Nature Switzerland}, \bibinfo{address}{Cham}, \bibinfo{year}{2026}).

\bibitem{gonsalves2019}
\bibinfo{author}{Gonsalves, A.~J.} \emph{et~al.}
\newblock \bibinfo{journal}{\bibinfo{title}{Petawatt {{Laser Guiding}} and {{Electron Beam Acceleration}} to 8 {{GeV}} in a {{Laser-Heated Capillary Discharge Waveguide}}}}.
\newblock {\emph{\JournalTitle{Physical Review Letters}}} \textbf{\bibinfo{volume}{122}}, \bibinfo{pages}{084801}, \doiprefix\url{10.1103/PhysRevLett.122.084801} (\bibinfo{year}{2019}).

\bibitem{aniculaesei2023}
\bibinfo{author}{Aniculaesei, C.} \emph{et~al.}
\newblock \bibinfo{journal}{\bibinfo{title}{The acceleration of a high-charge electron bunch to 10 {{GeV}} in a 10-cm nanoparticle-assisted wakefield accelerator}}.
\newblock {\emph{\JournalTitle{Matter and Radiation at Extremes}}} \textbf{\bibinfo{volume}{9}}, \bibinfo{pages}{014001}, \doiprefix\url{10.1063/5.0161687} (\bibinfo{year}{2023}).

\bibitem{picksley2024}
\bibinfo{author}{Picksley, A.} \emph{et~al.}
\newblock \bibinfo{journal}{\bibinfo{title}{Matched {{Guiding}} and {{Controlled Injection}} in {{Dark-Current-Free}}, 10-{{GeV-Class}}, {{Channel-Guided Laser-Plasma Accelerators}}}}.
\newblock {\emph{\JournalTitle{Physical Review Letters}}} \textbf{\bibinfo{volume}{133}}, \bibinfo{pages}{255001}, \doiprefix\url{10.1103/PhysRevLett.133.255001} (\bibinfo{year}{2024}).

\bibitem{couperus2017}
\bibinfo{author}{Couperus, J.~P.} \emph{et~al.}
\newblock \bibinfo{journal}{\bibinfo{title}{Demonstration of a beam loaded nanocoulomb-class laser wakefield accelerator}}.
\newblock {\emph{\JournalTitle{Nature Communications}}} \textbf{\bibinfo{volume}{8}}, \bibinfo{pages}{487}, \doiprefix\url{10.1038/s41467-017-00592-7} (\bibinfo{year}{2017}).

\bibitem{rockafellow2025}
\bibinfo{author}{Rockafellow, E.} \emph{et~al.}
\newblock \bibinfo{journal}{\bibinfo{title}{Development of a high charge 10 {{GeV}} laser electron accelerator}}.
\newblock {\emph{\JournalTitle{Physics of Plasmas}}} \textbf{\bibinfo{volume}{32}}, \bibinfo{pages}{053102}, \doiprefix\url{10.1063/5.0265640} (\bibinfo{year}{2025}).

\bibitem{guenot2017}
\bibinfo{author}{Gu{\'e}not, D.} \emph{et~al.}
\newblock \bibinfo{journal}{\bibinfo{title}{Relativistic electron beams driven by {{kHz}} single-cycle light pulses}}.
\newblock {\emph{\JournalTitle{Nature Photonics}}} \textbf{\bibinfo{volume}{11}}, \bibinfo{pages}{293--296}, \doiprefix\url{10.1038/nphoton.2017.46} (\bibinfo{year}{2017}).

\bibitem{faure2018}
\bibinfo{author}{Faure, J.} \emph{et~al.}
\newblock \bibinfo{journal}{\bibinfo{title}{A review of recent progress on laser-plasma acceleration at {{kHz}} repetition rate}}.
\newblock {\emph{\JournalTitle{Plasma Physics and Controlled Fusion}}} \textbf{\bibinfo{volume}{61}}, \bibinfo{pages}{014012}, \doiprefix\url{10.1088/1361-6587/aae047} (\bibinfo{year}{2018}).

\bibitem{salehi2021}
\bibinfo{author}{Salehi, F.}, \bibinfo{author}{Le, M.}, \bibinfo{author}{Railing, L.}, \bibinfo{author}{Kolesik, M.} \& \bibinfo{author}{Milchberg, H.~M.}
\newblock \bibinfo{journal}{\bibinfo{title}{Laser-{{Accelerated}}, {{Low-Divergence}} 15-{{MeV Quasimonoenergetic Electron Bunches}} at 1 {{kHz}}}}.
\newblock {\emph{\JournalTitle{Physical Review X}}} \textbf{\bibinfo{volume}{11}}, \bibinfo{pages}{021055}, \doiprefix\url{10.1103/PhysRevX.11.021055} (\bibinfo{year}{2021}).

\bibitem{lazzarini2024}
\bibinfo{author}{Lazzarini, C.~M.} \emph{et~al.}
\newblock \bibinfo{journal}{\bibinfo{title}{Ultrarelativistic electron beams accelerated by terawatt scalable {{kHz}} laser}}.
\newblock {\emph{\JournalTitle{Physics of Plasmas}}} \textbf{\bibinfo{volume}{31}}, \bibinfo{pages}{030703}, \doiprefix\url{10.1063/5.0189051} (\bibinfo{year}{2024}).

\bibitem{gruner2007}
\bibinfo{author}{Gr{\"u}ner, F.} \emph{et~al.}
\newblock \bibinfo{journal}{\bibinfo{title}{Design considerations for table-top, laser-based {{VUV}} and {{X-ray}} free electron lasers}}.
\newblock {\emph{\JournalTitle{Applied Physics B}}} \textbf{\bibinfo{volume}{86}}, \bibinfo{pages}{431--435}, \doiprefix\url{10.1007/s00340-006-2565-7} (\bibinfo{year}{2007}).

\bibitem{corde2013}
\bibinfo{author}{Corde, S.} \emph{et~al.}
\newblock \bibinfo{journal}{\bibinfo{title}{Femtosecond x rays from laser-plasma accelerators}}.
\newblock {\emph{\JournalTitle{Reviews of Modern Physics}}} \textbf{\bibinfo{volume}{85}}, \bibinfo{pages}{1--48}, \doiprefix\url{10.1103/RevModPhys.85.1} (\bibinfo{year}{2013}).

\bibitem{bulanov2013}
\bibinfo{author}{Bulanov, S.~V.}, \bibinfo{author}{Esirkepov, T.~{\relax Zh}.}, \bibinfo{author}{Kando, M.}, \bibinfo{author}{Pirozhkov, A.~S.} \& \bibinfo{author}{Rosanov, N.~N.}
\newblock \bibinfo{journal}{\bibinfo{title}{Relativistic mirrors in plasmas. {{Novel}} results and perspectives}}.
\newblock {\emph{\JournalTitle{Physics-Uspekhi}}} \textbf{\bibinfo{volume}{56}}, \bibinfo{pages}{429--464}, \doiprefix\url{10.3367/ufne.0183.201305a.0449} (\bibinfo{year}{2013}).

\bibitem{albert2014}
\bibinfo{author}{Albert, F.} \emph{et~al.}
\newblock \bibinfo{journal}{\bibinfo{title}{Laser wakefield accelerator based light sources: {{Potential}} applications and requirements}}.
\newblock {\emph{\JournalTitle{Plasma Physics and Controlled Fusion}}} \textbf{\bibinfo{volume}{56}}, \bibinfo{pages}{084015}, \doiprefix\url{10.1088/0741-3335/56/8/084015} (\bibinfo{year}{2014}).

\bibitem{kurz2021}
\bibinfo{author}{Kurz, T.} \emph{et~al.}
\newblock \bibinfo{journal}{\bibinfo{title}{Demonstration of a compact plasma accelerator powered by laser-accelerated electron beams}}.
\newblock {\emph{\JournalTitle{Nature Communications}}} \textbf{\bibinfo{volume}{12}}, \bibinfo{pages}{2895}, \doiprefix\url{10.1038/s41467-021-23000-7} (\bibinfo{year}{2021}).

\bibitem{williams2015}
\bibinfo{author}{Williams, G.~J.}, \bibinfo{author}{Pollock, B.~B.}, \bibinfo{author}{Albert, F.}, \bibinfo{author}{Park, J.} \& \bibinfo{author}{Chen, H.}
\newblock \bibinfo{journal}{\bibinfo{title}{Positron generation using laser-wakefield electron sources}}.
\newblock {\emph{\JournalTitle{Physics of Plasmas}}} \textbf{\bibinfo{volume}{22}}, \bibinfo{pages}{093115}, \doiprefix\url{10.1063/1.4931044} (\bibinfo{year}{2015}).

\bibitem{streeter2024}
\bibinfo{author}{Streeter, M. J.~V.} \emph{et~al.}
\newblock \bibinfo{journal}{\bibinfo{title}{Narrow bandwidth, low-emittance positron beams from a laser-wakefield accelerator}}.
\newblock {\emph{\JournalTitle{Scientific Reports}}} \textbf{\bibinfo{volume}{14}}, \bibinfo{pages}{6001}, \doiprefix\url{10.1038/s41598-024-56281-1} (\bibinfo{year}{2024}).

\bibitem{zhang2025}
\bibinfo{author}{Zhang, F.} \emph{et~al.}
\newblock \bibinfo{journal}{\bibinfo{title}{Proof-of-principle demonstration of muon production with an ultrashort high-intensity laser}}.
\newblock {\emph{\JournalTitle{Nature Physics}}} \textbf{\bibinfo{volume}{21}}, \bibinfo{pages}{1050--1056}, \doiprefix\url{10.1038/s41567-025-02872-2} (\bibinfo{year}{2025}).

\bibitem{ludwig2025}
\bibinfo{author}{Ludwig, J.~D.} \emph{et~al.}
\newblock \bibinfo{journal}{\bibinfo{title}{Laser based 100 {{GeV}} electron acceleration scheme for muon production}}.
\newblock {\emph{\JournalTitle{Scientific Reports}}} \textbf{\bibinfo{volume}{15}}, \bibinfo{pages}{25902}, \doiprefix\url{10.1038/s41598-025-95440-w} (\bibinfo{year}{2025}).

\bibitem{terzani2025}
\bibinfo{author}{Terzani, D.} \emph{et~al.}
\newblock \bibinfo{journal}{\bibinfo{title}{Measurement of directional muon beams generated at the berkeley lab laser accelerator}}.
\newblock {\emph{\JournalTitle{Physical Review Accelerators and Beams}}} \textbf{\bibinfo{volume}{28}}, \bibinfo{pages}{103401}, \doiprefix\url{10.1103/kxjr-h7zs} (\bibinfo{year}{2025}).

\bibitem{calvin2026}
\bibinfo{author}{Calvin, L.} \emph{et~al.}
\newblock \bibinfo{journal}{\bibinfo{title}{Experimental evidence of production of directional muons from a laser-wakefield accelerator}}.
\newblock {\emph{\JournalTitle{Plasma Physics and Controlled Fusion}}} \textbf{\bibinfo{volume}{68}}, \bibinfo{pages}{035015}, \doiprefix\url{10.1088/1361-6587/ae4d05} (\bibinfo{year}{2026}).

\bibitem{yan2017}
\bibinfo{author}{Yan, W.} \emph{et~al.}
\newblock \bibinfo{journal}{\bibinfo{title}{High-order multiphoton {{Thomson}} scattering}}.
\newblock {\emph{\JournalTitle{Nature Photonics}}} \textbf{\bibinfo{volume}{11}}, \bibinfo{pages}{514--520}, \doiprefix\url{10.1038/nphoton.2017.100} (\bibinfo{year}{2017}).

\bibitem{cole2018}
\bibinfo{author}{Cole, J.~M.} \emph{et~al.}
\newblock \bibinfo{journal}{\bibinfo{title}{Experimental {{Evidence}} of {{Radiation Reaction}} in the {{Collision}} of a {{High-Intensity Laser Pulse}} with a {{Laser-Wakefield Accelerated Electron Beam}}}}.
\newblock {\emph{\JournalTitle{Physical Review X}}} \textbf{\bibinfo{volume}{8}}, \bibinfo{pages}{011020}, \doiprefix\url{10.1103/PhysRevX.8.011020} (\bibinfo{year}{2018}).

\bibitem{poder2018}
\bibinfo{author}{Poder, K.} \emph{et~al.}
\newblock \bibinfo{journal}{\bibinfo{title}{Experimental {{Signatures}} of the {{Quantum Nature}} of {{Radiation Reaction}} in the {{Field}} of an {{Ultraintense Laser}}}}.
\newblock {\emph{\JournalTitle{Physical Review X}}} \textbf{\bibinfo{volume}{8}}, \bibinfo{pages}{031004}, \doiprefix\url{10.1103/PhysRevX.8.031004} (\bibinfo{year}{2018}).

\bibitem{gonoskov2022}
\bibinfo{author}{Gonoskov, A.}, \bibinfo{author}{Blackburn, T.~G.}, \bibinfo{author}{Marklund, M.} \& \bibinfo{author}{Bulanov, S.~S.}
\newblock \bibinfo{journal}{\bibinfo{title}{Charged particle motion and radiation in strong electromagnetic fields}}.
\newblock {\emph{\JournalTitle{Reviews of Modern Physics}}} \textbf{\bibinfo{volume}{94}}, \bibinfo{pages}{045001}, \doiprefix\url{10.1103/RevModPhys.94.045001} (\bibinfo{year}{2022}).

\bibitem{russell2023}
\bibinfo{author}{Russell, B.~K.} \emph{et~al.}
\newblock \bibinfo{journal}{\bibinfo{title}{Ultrafast relativistic electron probing of extreme magnetic fields}}.
\newblock {\emph{\JournalTitle{Physics of Plasmas}}} \textbf{\bibinfo{volume}{30}}, \bibinfo{pages}{093105}, \doiprefix\url{10.1063/5.0163392} (\bibinfo{year}{2023}).

\bibitem{mirzaie2024}
\bibinfo{author}{Mirzaie, M.} \emph{et~al.}
\newblock \bibinfo{journal}{\bibinfo{title}{All-optical nonlinear {{Compton}} scattering performed with a multi-petawatt laser}}.
\newblock {\emph{\JournalTitle{Nature Photonics}}} \textbf{\bibinfo{volume}{18}}, \bibinfo{pages}{1212--1217}, \doiprefix\url{10.1038/s41566-024-01550-8} (\bibinfo{year}{2024}).

\bibitem{russell2024}
\bibinfo{author}{Russell, B.~K.} \emph{et~al.}
\newblock \bibinfo{journal}{\bibinfo{title}{Measuring signatures in photon angular spectra to distinguish nonlinear {{Compton}} scattering models}}.
\newblock {\emph{\JournalTitle{Physical Review A}}} \textbf{\bibinfo{volume}{110}}, \bibinfo{pages}{062822}, \doiprefix\url{10.1103/PhysRevA.110.062822} (\bibinfo{year}{2024}).

\bibitem{shalloo2020}
\bibinfo{author}{Shalloo, R.~J.} \emph{et~al.}
\newblock \bibinfo{journal}{\bibinfo{title}{Automation and control of laser wakefield accelerators using {{Bayesian}} optimization}}.
\newblock {\emph{\JournalTitle{Nature Communications}}} \textbf{\bibinfo{volume}{11}}, \bibinfo{pages}{6355}, \doiprefix\url{10.1038/s41467-020-20245-6} (\bibinfo{year}{2020}).

\bibitem{jalas2021}
\bibinfo{author}{Jalas, S.} \emph{et~al.}
\newblock \bibinfo{journal}{\bibinfo{title}{Bayesian {{Optimization}} of a {{Laser-Plasma Accelerator}}}}.
\newblock {\emph{\JournalTitle{Physical Review Letters}}} \textbf{\bibinfo{volume}{126}}, \bibinfo{pages}{104801}, \doiprefix\url{10.1103/PhysRevLett.126.104801} (\bibinfo{year}{2021}).

\bibitem{kirchen2021}
\bibinfo{author}{Kirchen, M.} \emph{et~al.}
\newblock \bibinfo{journal}{\bibinfo{title}{Optimal {{Beam Loading}} in a {{Laser-Plasma Accelerator}}}}.
\newblock {\emph{\JournalTitle{Physical Review Letters}}} \textbf{\bibinfo{volume}{126}}, \bibinfo{pages}{174801}, \doiprefix\url{10.1103/PhysRevLett.126.174801} (\bibinfo{year}{2021}).

\bibitem{dopp2023a}
\bibinfo{author}{D{\"o}pp, A.} \emph{et~al.}
\newblock \bibinfo{journal}{\bibinfo{title}{Data-driven science and machine learning methods in laser--plasma physics}}.
\newblock {\emph{\JournalTitle{High Power Laser Science and Engineering}}} \textbf{\bibinfo{volume}{11}}, \bibinfo{pages}{e55}, \doiprefix\url{10.1017/hpl.2023.47} (\bibinfo{year}{2023}).

\bibitem{roussel2024}
\bibinfo{author}{Roussel, R.} \emph{et~al.}
\newblock \bibinfo{journal}{\bibinfo{title}{Bayesian optimization algorithms for accelerator physics}}.
\newblock {\emph{\JournalTitle{Physical Review Accelerators and Beams}}} \textbf{\bibinfo{volume}{27}}, \bibinfo{pages}{084801}, \doiprefix\url{10.1103/PhysRevAccelBeams.27.084801} (\bibinfo{year}{2024}).

\bibitem{irshad2024}
\bibinfo{author}{Irshad, F.} \emph{et~al.}
\newblock \bibinfo{journal}{\bibinfo{title}{Pareto {{Optimization}} and {{Tuning}} of a {{Laser Wakefield Accelerator}}}}.
\newblock {\emph{\JournalTitle{Physical Review Letters}}} \textbf{\bibinfo{volume}{133}}, \bibinfo{pages}{085001}, \doiprefix\url{10.1103/PhysRevLett.133.085001} (\bibinfo{year}{2024}).

\bibitem{valenta2025}
\bibinfo{author}{Valenta, P.}, \bibinfo{author}{Esirkepov, T.~{\relax Zh}.}, \bibinfo{author}{Ludwig, J.~D.}, \bibinfo{author}{Wilks, S.~C.} \& \bibinfo{author}{Bulanov, S.~V.}
\newblock \bibinfo{journal}{\bibinfo{title}{Bayesian optimization of electron energy from laser wakefield accelerators}}.
\newblock {\emph{\JournalTitle{Physical Review Accelerators and Beams}}} \textbf{\bibinfo{volume}{28}}, \bibinfo{pages}{094601}, \doiprefix\url{10.1103/knh7-hbr3} (\bibinfo{year}{2025}).

\bibitem{valenta2026}
\bibinfo{author}{Valenta, P.} \emph{et~al.}
\newblock \bibinfo{journal}{\bibinfo{title}{Optimized matching conditions for self-guided laser wakefield accelerators}}.
\newblock {\emph{\JournalTitle{Machine Learning: Science and Technology}}} \textbf{\bibinfo{volume}{7}}, \bibinfo{pages}{025030}, \doiprefix\url{10.1088/2632-2153/ae51e0} (\bibinfo{year}{2026}).

\bibitem{maslarova2026}
\bibinfo{author}{Maslarova, D.} \emph{et~al.}
\newblock \bibinfo{journal}{\bibinfo{title}{Batch {{Bayesian}} optimization of attosecond betatron pulses from laser wakefield acceleration}}.
\newblock {\emph{\JournalTitle{Communications Physics}}} \textbf{\bibinfo{volume}{9}}, \bibinfo{pages}{92}, \doiprefix\url{10.1038/s42005-026-02542-6} (\bibinfo{year}{2026}).

\bibitem{sun1987}
\bibinfo{author}{Sun, G.-Z.}, \bibinfo{author}{Ott, E.}, \bibinfo{author}{Lee, Y.~C.} \& \bibinfo{author}{Guzdar, P.}
\newblock \bibinfo{journal}{\bibinfo{title}{Self-focusing of short intense pulses in plasmas}}.
\newblock {\emph{\JournalTitle{Physics of Fluids}}} \textbf{\bibinfo{volume}{30}}, \bibinfo{pages}{526}, \doiprefix\url{10.1063/1.866349} (\bibinfo{year}{1987}).

\bibitem{naumova2002}
\bibinfo{author}{Naumova, N.~M.}, \bibinfo{author}{Bulanov, S.~V.}, \bibinfo{author}{Nishihara, K.}, \bibinfo{author}{Esirkepov, T.~{\relax Zh}.} \& \bibinfo{author}{Pegoraro, F.}
\newblock \bibinfo{journal}{\bibinfo{title}{Polarization effects and anisotropy in three-dimensional relativistic self-focusing}}.
\newblock {\emph{\JournalTitle{Physical Review E}}} \textbf{\bibinfo{volume}{65}}, \bibinfo{pages}{045402}, \doiprefix\url{10.1103/PhysRevE.65.045402} (\bibinfo{year}{2002}).

\bibitem{valenta2021}
\bibinfo{author}{Valenta, P.}, \bibinfo{author}{Grittani, G.~M.}, \bibinfo{author}{Lazzarini, C.~M.}, \bibinfo{author}{Klimo, O.} \& \bibinfo{author}{Bulanov, S.~V.}
\newblock \bibinfo{journal}{\bibinfo{title}{On the electromagnetic-electron rings originating from the interaction of high-power short-pulse laser and underdense plasma}}.
\newblock {\emph{\JournalTitle{Physics of Plasmas}}} \textbf{\bibinfo{volume}{28}}, \bibinfo{pages}{122104}, \doiprefix\url{10.1063/5.0065167} (\bibinfo{year}{2021}).

\bibitem{bulanov1996}
\bibinfo{author}{Bulanov, S.~V.}, \bibinfo{author}{Lontano, M.}, \bibinfo{author}{Esirkepov, T.~{\relax Zh}.}, \bibinfo{author}{Pegoraro, F.} \& \bibinfo{author}{Pukhov, A.~M.}
\newblock \bibinfo{journal}{\bibinfo{title}{Electron vortices produced by ultraintense laser pulses}}.
\newblock {\emph{\JournalTitle{Physical Review Letters}}} \textbf{\bibinfo{volume}{76}}, \bibinfo{pages}{3562--3565}, \doiprefix\url{10.1103/PhysRevLett.76.3562} (\bibinfo{year}{1996}).

\bibitem{bulanov1999}
\bibinfo{author}{Bulanov, S.~V.}, \bibinfo{author}{Esirkepov, T.~{\relax Zh}.}, \bibinfo{author}{Naumova, N.~M.}, \bibinfo{author}{Pegoraro, F.} \& \bibinfo{author}{Vshivkov, V.~A.}
\newblock \bibinfo{journal}{\bibinfo{title}{Solitonlike electromagnetic waves behind a superintense laser pulse in a plasma}}.
\newblock {\emph{\JournalTitle{Physical Review Letters}}} \textbf{\bibinfo{volume}{82}}, \bibinfo{pages}{3440--3443}, \doiprefix\url{10.1103/PhysRevLett.82.3440} (\bibinfo{year}{1999}).

\bibitem{esirkepov2002}
\bibinfo{author}{Esirkepov, T.~{\relax Zh}.}, \bibinfo{author}{Nishihara, K.}, \bibinfo{author}{Bulanov, S.~V.} \& \bibinfo{author}{Pegoraro, F.}
\newblock \bibinfo{journal}{\bibinfo{title}{Three-dimensional relativistic electromagnetic subcycle solitons}}.
\newblock {\emph{\JournalTitle{Physical Review Letters}}} \textbf{\bibinfo{volume}{89}}, \bibinfo{pages}{275002}, \doiprefix\url{10.1103/PhysRevLett.89.275002} (\bibinfo{year}{2002}).

\bibitem{lu2006}
\bibinfo{author}{Lu, W.}, \bibinfo{author}{Huang, C.}, \bibinfo{author}{Zhou, M.}, \bibinfo{author}{Mori, W.~B.} \& \bibinfo{author}{Katsouleas, T.}
\newblock \bibinfo{journal}{\bibinfo{title}{Nonlinear theory for relativistic plasma wakefields in the blowout regime}}.
\newblock {\emph{\JournalTitle{Physical Review Letters}}} \textbf{\bibinfo{volume}{96}}, \bibinfo{pages}{165002}, \doiprefix\url{10.1103/PhysRevLett.96.165002} (\bibinfo{year}{2006}).

\bibitem{lu2007}
\bibinfo{author}{Lu, W.} \emph{et~al.}
\newblock \bibinfo{journal}{\bibinfo{title}{Generating multi-{{GeV}} electron bunches using single stage laser wakefield acceleration in a {{3D}} nonlinear regime}}.
\newblock {\emph{\JournalTitle{Physical Review Special Topics - Accelerators and Beams}}} \textbf{\bibinfo{volume}{10}}, \bibinfo{pages}{61301}, \doiprefix\url{10.1103/PhysRevSTAB.10.061301} (\bibinfo{year}{2007}).

\bibitem{kostyukov2004}
\bibinfo{author}{Kostyukov, I.}, \bibinfo{author}{Pukhov, A.} \& \bibinfo{author}{Kiselev, S.}
\newblock \bibinfo{journal}{\bibinfo{title}{Phenomenological theory of laser-plasma interaction in ``bubble'' regime}}.
\newblock {\emph{\JournalTitle{Physics of Plasmas}}} \textbf{\bibinfo{volume}{11}}, \bibinfo{pages}{5256--5264}, \doiprefix\url{10.1063/1.1799371} (\bibinfo{year}{2004}).

\bibitem{gordienko2005}
\bibinfo{author}{Gordienko, S.} \& \bibinfo{author}{Pukhov, A.}
\newblock \bibinfo{journal}{\bibinfo{title}{Scalings for ultrarelativistic laser plasmas and quasimonoenergetic electrons}}.
\newblock {\emph{\JournalTitle{Physics of Plasmas}}} \textbf{\bibinfo{volume}{12}}, \bibinfo{pages}{043109}, \doiprefix\url{10.1063/1.1884126} (\bibinfo{year}{2005}).

\bibitem{tzoufras2009}
\bibinfo{author}{Tzoufras, M.} \emph{et~al.}
\newblock \bibinfo{journal}{\bibinfo{title}{Beam loading by electrons in nonlinear plasma wakes}}.
\newblock {\emph{\JournalTitle{Physics of Plasmas}}} \textbf{\bibinfo{volume}{16}}, \bibinfo{pages}{056705}, \doiprefix\url{10.1063/1.3118628} (\bibinfo{year}{2009}).

\bibitem{schroeder2010}
\bibinfo{author}{Schroeder, C.~B.}, \bibinfo{author}{Esarey, E.}, \bibinfo{author}{Geddes, C. G.~R.}, \bibinfo{author}{Benedetti, C.} \& \bibinfo{author}{Leemans, W.~P.}
\newblock \bibinfo{journal}{\bibinfo{title}{Physics considerations for laser-plasma linear colliders}}.
\newblock {\emph{\JournalTitle{Physical Review Special Topics - Accelerators and Beams}}} \textbf{\bibinfo{volume}{13}}, \bibinfo{pages}{101301}, \doiprefix\url{10.1103/PhysRevSTAB.13.101301} (\bibinfo{year}{2010}).

\bibitem{yi2013}
\bibinfo{author}{Yi, S.~A.}, \bibinfo{author}{Khudik, V.}, \bibinfo{author}{Siemon, C.} \& \bibinfo{author}{Shvets, G.}
\newblock \bibinfo{journal}{\bibinfo{title}{Analytic model of electromagnetic fields around a plasma bubble in the blow-out regime}}.
\newblock {\emph{\JournalTitle{Physics of Plasmas}}} \textbf{\bibinfo{volume}{20}}, \bibinfo{pages}{013108}, \doiprefix\url{10.1063/1.4775774} (\bibinfo{year}{2013}).

\bibitem{davidson2019}
\bibinfo{author}{Davidson, A.} \emph{et~al.}
\newblock \bibinfo{title}{Optimizing {{Laser Wakefield Acceleration}} in the {{Nonlinear Self-Guided Regime}} for {{Fixed Laser Energy}}}, \doiprefix\url{10.48550/arXiv.1805.08761} (\bibinfo{year}{2019}).
\newblock \eprint{1805.08761}.

\bibitem{dalichaouch2021}
\bibinfo{author}{Dalichaouch, T.~N.} \emph{et~al.}
\newblock \bibinfo{journal}{\bibinfo{title}{A multi-sheath model for highly nonlinear plasma wakefields}}.
\newblock {\emph{\JournalTitle{Physics of Plasmas}}} \textbf{\bibinfo{volume}{28}}, \bibinfo{pages}{063103}, \doiprefix\url{10.1063/5.0051282} (\bibinfo{year}{2021}).

\bibitem{golovanov2023}
\bibinfo{author}{Golovanov, A.}, \bibinfo{author}{Kostyukov, I.~{\relax Yu}.}, \bibinfo{author}{Pukhov, A.} \& \bibinfo{author}{Malka, V.}
\newblock \bibinfo{journal}{\bibinfo{title}{Energy-{{Conserving Theory}} of the {{Blowout Regime}} of {{Plasma Wakefield}}}}.
\newblock {\emph{\JournalTitle{Physical Review Letters}}} \textbf{\bibinfo{volume}{130}}, \bibinfo{pages}{105001}, \doiprefix\url{10.1103/PhysRevLett.130.105001} (\bibinfo{year}{2023}).

\bibitem{lifschitz2009}
\bibinfo{author}{Lifschitz, A.~F.} \emph{et~al.}
\newblock \bibinfo{journal}{\bibinfo{title}{Particle-in-{{Cell}} modelling of laser--plasma interaction using {{Fourier}} decomposition}}.
\newblock {\emph{\JournalTitle{Journal of Computational Physics}}} \textbf{\bibinfo{volume}{228}}, \bibinfo{pages}{1803--1814}, \doiprefix\url{10.1016/j.jcp.2008.11.017} (\bibinfo{year}{2009}).

\bibitem{davidson2015}
\bibinfo{author}{Davidson, A.} \emph{et~al.}
\newblock \bibinfo{journal}{\bibinfo{title}{Implementation of a hybrid particle code with a {{PIC}} description in r--z and a gridless description in {{$\phi$}} into {{OSIRIS}}}}.
\newblock {\emph{\JournalTitle{Journal of Computational Physics}}} \textbf{\bibinfo{volume}{281}}, \bibinfo{pages}{1063--1077}, \doiprefix\url{10.1016/j.jcp.2014.10.064} (\bibinfo{year}{2015}).

\bibitem{vay2007}
\bibinfo{author}{Vay, J.-L.}
\newblock \bibinfo{journal}{\bibinfo{title}{Noninvariance of {{Space-}} and {{Time-Scale Ranges}} under a {{Lorentz Transformation}} and the {{Implications}} for the {{Study}} of {{Relativistic Interactions}}}}.
\newblock {\emph{\JournalTitle{Physical Review Letters}}} \textbf{\bibinfo{volume}{98}}, \bibinfo{pages}{130405}, \doiprefix\url{10.1103/PhysRevLett.98.130405} (\bibinfo{year}{2007}).

\bibitem{yu2016}
\bibinfo{author}{Yu, P.} \emph{et~al.}
\newblock \bibinfo{journal}{\bibinfo{title}{Enabling {{Lorentz}} boosted frame particle-in-cell simulations of laser wakefield acceleration in quasi-{{3D}} geometry}}.
\newblock {\emph{\JournalTitle{Journal of Computational Physics}}} \textbf{\bibinfo{volume}{316}}, \bibinfo{pages}{747--759}, \doiprefix\url{10.1016/j.jcp.2016.04.014} (\bibinfo{year}{2016}).

\bibitem{nedorezov2021}
\bibinfo{author}{Nedorezov, V.~G.}, \bibinfo{author}{Rykovanov, S.~G.} \& \bibinfo{author}{Savel'ev, A.~B.}
\newblock \bibinfo{journal}{\bibinfo{title}{Nuclear photonics: {{Results}} and prospects}}.
\newblock {\emph{\JournalTitle{Physics-Uspekhi}}} \textbf{\bibinfo{volume}{64}}, \bibinfo{pages}{1214--1237} (\bibinfo{year}{2021}).

\bibitem{kolenaty2022}
\bibinfo{author}{Kolenat{\'y}, D.} \emph{et~al.}
\newblock \bibinfo{journal}{\bibinfo{title}{Electron-positron pairs and radioactive nuclei production by irradiation of high-{{Z}} target with gamma-photon flash generated by an ultra-intense laser in the lambda\textasciicircum 3 regime}}.
\newblock {\emph{\JournalTitle{Physical Review Research}}} \textbf{\bibinfo{volume}{4}}, \bibinfo{pages}{023124}, \doiprefix\url{10.1103/PhysRevResearch.4.023124} (\bibinfo{year}{2022}).

\bibitem{beaurepaire2015}
\bibinfo{author}{Beaurepaire, B.} \emph{et~al.}
\newblock \bibinfo{journal}{\bibinfo{title}{Effect of the {{Laser Wave Front}} in a {{Laser-Plasma Accelerator}}}}.
\newblock {\emph{\JournalTitle{Physical Review X}}} \textbf{\bibinfo{volume}{5}}, \bibinfo{pages}{031012}, \doiprefix\url{10.1103/PhysRevX.5.031012} (\bibinfo{year}{2015}).

\bibitem{oumbarekespinos2023}
\bibinfo{author}{Oumbarek~Espinos, D.} \emph{et~al.}
\newblock \bibinfo{journal}{\bibinfo{title}{Notable improvements on {{LWFA}} through precise laser wavefront tuning}}.
\newblock {\emph{\JournalTitle{Scientific Reports}}} \textbf{\bibinfo{volume}{13}}, \bibinfo{pages}{18466}, \doiprefix\url{10.1038/s41598-023-45737-5} (\bibinfo{year}{2023}).

\bibitem{kalmykov2012}
\bibinfo{author}{Kalmykov, S.~Y.}, \bibinfo{author}{Beck, A.}, \bibinfo{author}{Davoine, X.}, \bibinfo{author}{Lefebvre, E.} \& \bibinfo{author}{Shadwick, B.~A.}
\newblock \bibinfo{journal}{\bibinfo{title}{Laser plasma acceleration with a negatively chirped pulse: {{All-optical}} control over dark current in the blowout regime}}.
\newblock {\emph{\JournalTitle{New Journal of Physics}}} \textbf{\bibinfo{volume}{14}}, \bibinfo{pages}{033025}, \doiprefix\url{10.1088/1367-2630/14/3/033025} (\bibinfo{year}{2012}).

\bibitem{kim2017a}
\bibinfo{author}{Kim, H.~T.} \emph{et~al.}
\newblock \bibinfo{journal}{\bibinfo{title}{Stable multi-{{GeV}} electron accelerator driven by waveform-controlled {{PW}} laser pulses}}.
\newblock {\emph{\JournalTitle{Scientific Reports}}} \textbf{\bibinfo{volume}{7}}, \bibinfo{pages}{10203}, \doiprefix\url{10.1038/s41598-017-09267-1} (\bibinfo{year}{2017}).

\bibitem{bulanov1993}
\bibinfo{author}{Bulanov, S.~V.}, \bibinfo{author}{Kirsanov, V.~I.}, \bibinfo{author}{Pegoraro, F.} \& \bibinfo{author}{Sakharov, A.~S.}
\newblock \bibinfo{journal}{\bibinfo{title}{Charged particle and photon acceleration by wake field plasma waves in nonuniform plasmas}}.
\newblock {\emph{\JournalTitle{Laser Physics}}} \textbf{\bibinfo{volume}{3}}, \bibinfo{pages}{1078} (\bibinfo{year}{1993}).

\bibitem{bulanov1997}
\bibinfo{author}{Bulanov, S.~V.} \emph{et~al.}
\newblock \bibinfo{journal}{\bibinfo{title}{Laser acceleration of charged particles in inhomogeneous plasmas {{I}}.}}
\newblock {\emph{\JournalTitle{Plasma Physics Reports}}} \textbf{\bibinfo{volume}{23}}, \bibinfo{pages}{259--269} (\bibinfo{year}{1997}).

\bibitem{sprangle2001}
\bibinfo{author}{Sprangle, P.} \emph{et~al.}
\newblock \bibinfo{journal}{\bibinfo{title}{Wakefield generation and {{GeV}} acceleration in tapered plasma channels}}.
\newblock {\emph{\JournalTitle{Physical Review E}}} \textbf{\bibinfo{volume}{63}}, \bibinfo{pages}{056405}, \doiprefix\url{10.1103/PhysRevE.63.056405} (\bibinfo{year}{2001}).

\bibitem{guillaume2015}
\bibinfo{author}{Guillaume, E.} \emph{et~al.}
\newblock \bibinfo{journal}{\bibinfo{title}{Electron {{Rephasing}} in a {{Laser-Wakefield Accelerator}}}}.
\newblock {\emph{\JournalTitle{Physical Review Letters}}} \textbf{\bibinfo{volume}{115}}, \bibinfo{pages}{155002}, \doiprefix\url{10.1103/PhysRevLett.115.155002} (\bibinfo{year}{2015}).

\bibitem{steinke2016}
\bibinfo{author}{Steinke, S.} \emph{et~al.}
\newblock \bibinfo{journal}{\bibinfo{title}{Multistage coupling of independent laser-plasma accelerators}}.
\newblock {\emph{\JournalTitle{Nature}}} \textbf{\bibinfo{volume}{530}}, \bibinfo{pages}{190--193}, \doiprefix\url{10.1038/nature16525} (\bibinfo{year}{2016}).

\bibitem{luo2018}
\bibinfo{author}{Luo, J.} \emph{et~al.}
\newblock \bibinfo{journal}{\bibinfo{title}{Multistage {{Coupling}} of {{Laser-Wakefield Accelerators}} with {{Curved Plasma Channels}}}}.
\newblock {\emph{\JournalTitle{Physical Review Letters}}} \textbf{\bibinfo{volume}{120}}, \bibinfo{pages}{154801}, \doiprefix\url{10.1103/PhysRevLett.120.154801} (\bibinfo{year}{2018}).

\bibitem{haq2025}
\bibinfo{author}{Haq, R.~U.} \emph{et~al.}
\newblock \bibinfo{journal}{\bibinfo{title}{Multi-{{GeV}} electron beam generation via two-stage laser wakefield acceleration}}.
\newblock {\emph{\JournalTitle{Scientific Reports}}} \textbf{\bibinfo{volume}{15}}, \bibinfo{pages}{42290}, \doiprefix\url{10.1038/s41598-025-22766-w} (\bibinfo{year}{2025}).

\bibitem{esirkepov2006}
\bibinfo{author}{Esirkepov, T.}, \bibinfo{author}{Bulanov, S.~V.}, \bibinfo{author}{Yamagiwa, M.} \& \bibinfo{author}{Tajima, T.}
\newblock \bibinfo{journal}{\bibinfo{title}{Electron, {{Positron}}, and {{Photon Wakefield Acceleration}}: {{Trapping}}, {{Wake Overtaking}}, and {{Ponderomotive Acceleration}}}}.
\newblock {\emph{\JournalTitle{Physical Review Letters}}} \textbf{\bibinfo{volume}{96}}, \bibinfo{pages}{014803}, \doiprefix\url{10.1103/PhysRevLett.96.014803} (\bibinfo{year}{2006}).

\bibitem{leemans1996}
\bibinfo{author}{Leemans, W.~P.} \emph{et~al.}
\newblock \bibinfo{journal}{\bibinfo{title}{Plasma guiding and wakefield generation for second-generation experiments}}.
\newblock {\emph{\JournalTitle{IEEE Transactions on Plasma Science}}} \textbf{\bibinfo{volume}{24}}, \bibinfo{pages}{331--342}, \doiprefix\url{10.1109/27.509997} (\bibinfo{year}{1996}).

\bibitem{nerush2009}
\bibinfo{author}{Nerush, E.~N.} \& \bibinfo{author}{Kostyukov, I.~{\relax Yu}.}
\newblock \bibinfo{journal}{\bibinfo{title}{Carrier-envelope phase effects in plasma-based electron acceleration with few-cycle laser pulses}}.
\newblock {\emph{\JournalTitle{Physical Review Letters}}} \textbf{\bibinfo{volume}{103}}, \bibinfo{pages}{035001}, \doiprefix\url{10.1103/PhysRevLett.103.035001} (\bibinfo{year}{2009}).

\bibitem{valenta2020}
\bibinfo{author}{Valenta, P.} \emph{et~al.}
\newblock \bibinfo{journal}{\bibinfo{title}{Polarity reversal of wakefields driven by ultrashort pulse laser}}.
\newblock {\emph{\JournalTitle{Physical Review E}}} \textbf{\bibinfo{volume}{102}}, \bibinfo{pages}{53216}, \doiprefix\url{10.1103/PhysRevE.102.053216} (\bibinfo{year}{2020}).

\bibitem{huijts2021}
\bibinfo{author}{Huijts, J.}, \bibinfo{author}{Andriyash, I.~A.}, \bibinfo{author}{Rovige, L.}, \bibinfo{author}{Vernier, A.} \& \bibinfo{author}{Faure, J.}
\newblock \bibinfo{journal}{\bibinfo{title}{Identifying observable carrier-envelope phase effects in laser wakefield acceleration with near-{{Single-Cycle}} pulses}}.
\newblock {\emph{\JournalTitle{Physics of Plasmas}}} \textbf{\bibinfo{volume}{28}}, \bibinfo{pages}{043101}, \doiprefix\url{10.1063/5.0037925} (\bibinfo{year}{2021}).

\bibitem{wilks1987}
\bibinfo{author}{Wilks, S.}, \bibinfo{author}{Katsouleas, T.}, \bibinfo{author}{Dawson, J.~M.}, \bibinfo{author}{Chen, P.} \& \bibinfo{author}{Su, J.~J.}
\newblock \bibinfo{journal}{\bibinfo{title}{Beam {{Loading}} in {{Plasma Waves}}}}.
\newblock {\emph{\JournalTitle{IEEE Transactions on Plasma Science}}} \textbf{\bibinfo{volume}{15}}, \bibinfo{pages}{210--217}, \doiprefix\url{10.1109/TPS.1987.4316687} (\bibinfo{year}{1987}).

\bibitem{katsouleas1987}
\bibinfo{author}{Katsouleas, T.}, \bibinfo{author}{Wilks, S.}, \bibinfo{author}{Chen, P.}, \bibinfo{author}{Dawson, J.~M.} \& \bibinfo{author}{Su, J.~J.}
\newblock \bibinfo{journal}{\bibinfo{title}{Beam loading in plasma accelerators}}.
\newblock {\emph{\JournalTitle{Particle Accelerators}}} \textbf{\bibinfo{volume}{22}}, \bibinfo{pages}{81--99} (\bibinfo{year}{1987}).

\bibitem{fonseca2002}
\bibinfo{author}{Fonseca, R.~A.} \emph{et~al.}
\newblock \bibinfo{title}{{{OSIRIS}}: {{A}} three-dimensional, fully relativistic particle in cell code for modeling plasma based accelerators}.
\newblock Computational {{Science}} --- {{ICCS}} 2002, \bibinfo{pages}{342--351} (\bibinfo{publisher}{Springer Berlin Heidelberg}, \bibinfo{year}{2002}).

\bibitem{fonseca2008}
\bibinfo{author}{Fonseca, R.~A.} \emph{et~al.}
\newblock \bibinfo{journal}{\bibinfo{title}{One-to-one direct modeling of experiments and astrophysical scenarios: {{Pushing}} the envelope on kinetic plasma simulations}}.
\newblock {\emph{\JournalTitle{Plasma Physics and Controlled Fusion}}} \textbf{\bibinfo{volume}{50}}, \bibinfo{pages}{124034}, \doiprefix\url{10.1088/0741-3335/50/12/124034} (\bibinfo{year}{2008}).

\bibitem{martins2010}
\bibinfo{author}{Martins, S.~F.}, \bibinfo{author}{Fonseca, R.~A.}, \bibinfo{author}{Lu, W.}, \bibinfo{author}{Mori, W.~B.} \& \bibinfo{author}{Silva, L.~O.}
\newblock \bibinfo{journal}{\bibinfo{title}{Exploring laser-wakefield-accelerator regimes for near-{{Term}} lasers using particle-in-cell simulation in {{Lorentz-boosted}} frames}}.
\newblock {\emph{\JournalTitle{Nature Physics}}} \textbf{\bibinfo{volume}{6}}, \bibinfo{pages}{311--316}, \doiprefix\url{10.1038/nphys1538} (\bibinfo{year}{2010}).

\bibitem{yu2014}
\bibinfo{author}{Yu, P.} \emph{et~al.}
\newblock \bibinfo{journal}{\bibinfo{title}{Modeling of laser wakefield acceleration in {{Lorentz}} boosted frame using {{EM-PIC}} code with spectral solver}}.
\newblock {\emph{\JournalTitle{Journal of Computational Physics}}} \textbf{\bibinfo{volume}{266}}, \bibinfo{pages}{124--138}, \doiprefix\url{10.1016/j.jcp.2014.02.016} (\bibinfo{year}{2014}).

\bibitem{massimo2025}
\bibinfo{author}{Massimo, F.}, \bibinfo{author}{Benedetti, C.}, \bibinfo{author}{Terzani, D.}, \bibinfo{author}{Beck, A.} \& \bibinfo{author}{Cros, B.}
\newblock \bibinfo{journal}{\bibinfo{title}{Modeling laser-wakefield accelerators using the time-averaged ponderomotive approximation in a {{Lorentz}} boosted frame}}.
\newblock {\emph{\JournalTitle{Plasma Physics and Controlled Fusion}}} \textbf{\bibinfo{volume}{67}}, \bibinfo{pages}{065032}, \doiprefix\url{10.1088/1361-6587/addc97} (\bibinfo{year}{2025}).

\bibitem{boris1971}
\bibinfo{author}{Boris, J.~P.}
\newblock \bibinfo{title}{Relativistic plasma simulation - {{Optimization}} of a hybrid code}.
\newblock In \bibinfo{editor}{Boris, J.~P.} \& \bibinfo{editor}{Shanny, R.~A.} (eds.) \emph{\bibinfo{booktitle}{Proceedings of the Fourth Conference on Numerical Simulation of Plasmas}}, \bibinfo{pages}{3--67} (\bibinfo{publisher}{Naval Research Laboratory}, \bibinfo{year}{1971}).

\bibitem{li2017}
\bibinfo{author}{Li, F.} \emph{et~al.}
\newblock \bibinfo{journal}{\bibinfo{title}{Controlling the numerical {{Cerenkov}} instability in {{PIC}} simulations using a customized finite difference {{Maxwell}} solver and a local {{FFT}} based current correction}}.
\newblock {\emph{\JournalTitle{Computer Physics Communications}}} \textbf{\bibinfo{volume}{214}}, \bibinfo{pages}{6--17}, \doiprefix\url{10.1016/j.cpc.2017.01.001} (\bibinfo{year}{2017}).

\bibitem{hudson2022}
\bibinfo{author}{Hudson, S.}, \bibinfo{author}{Larson, J.}, \bibinfo{author}{Navarro, J.-L.} \& \bibinfo{author}{Wild, S.~M.}
\newblock \bibinfo{journal}{\bibinfo{title}{{{libEnsemble}}: {{A Library}} to {{Coordinate}} the {{Concurrent Evaluation}} of {{Dynamic Ensembles}} of {{Calculations}}}}.
\newblock {\emph{\JournalTitle{IEEE Transactions on Parallel and Distributed Systems}}} \textbf{\bibinfo{volume}{33}}, \bibinfo{pages}{977--988}, \doiprefix\url{10.1109/TPDS.2021.3082815} (\bibinfo{year}{2022}).

\bibitem{ferranpousa2023}
\bibinfo{author}{Ferran~Pousa, A.} \emph{et~al.}
\newblock \bibinfo{journal}{\bibinfo{title}{Bayesian optimization of laser-plasma accelerators assisted by reduced physical models}}.
\newblock {\emph{\JournalTitle{Physical Review Accelerators and Beams}}} \textbf{\bibinfo{volume}{26}}, \bibinfo{pages}{084601}, \doiprefix\url{10.1103/PhysRevAccelBeams.26.084601} (\bibinfo{year}{2023}).

\bibitem{jalas2023}
\bibinfo{author}{Jalas, S.} \emph{et~al.}
\newblock \bibinfo{journal}{\bibinfo{title}{Tuning curves for a laser-plasma accelerator}}.
\newblock {\emph{\JournalTitle{Physical Review Accelerators and Beams}}} \textbf{\bibinfo{volume}{26}}, \bibinfo{pages}{071302}, \doiprefix\url{10.1103/PhysRevAccelBeams.26.071302} (\bibinfo{year}{2023}).

\bibitem{irshad2023}
\bibinfo{author}{Irshad, F.}, \bibinfo{author}{Karsch, S.} \& \bibinfo{author}{D{\"o}pp, A.}
\newblock \bibinfo{journal}{\bibinfo{title}{Multi-objective and multi-fidelity {{Bayesian}} optimization of laser-plasma acceleration}}.
\newblock {\emph{\JournalTitle{Physical Review Research}}} \textbf{\bibinfo{volume}{5}}, \bibinfo{pages}{013063}, \doiprefix\url{10.1103/PhysRevResearch.5.013063} (\bibinfo{year}{2023}).

\bibitem{nunes2025}
\bibinfo{author}{Nunes, B.~S.} \emph{et~al.}
\newblock \bibinfo{journal}{\bibinfo{title}{Bayesian optimization of laser wakefield acceleration in the self-modulated regime ({{SM-LWFA}}) aiming to produce molybdenum-99 via photonuclear reactions}}.
\newblock {\emph{\JournalTitle{Physics of Plasmas}}} \textbf{\bibinfo{volume}{32}}, \bibinfo{pages}{033101}, \doiprefix\url{10.1063/5.0244268} (\bibinfo{year}{2025}).

\bibitem{djordjevic2026}
\bibinfo{author}{Djordjevi{\'c}, B.~Z.} \emph{et~al.}
\newblock \bibinfo{journal}{\bibinfo{title}{Bayesian optimization of laser wakefield acceleration via spectral pulse shaping}}.
\newblock {\emph{\JournalTitle{Physics of Plasmas}}} \textbf{\bibinfo{volume}{33}}, \bibinfo{pages}{063101}, \doiprefix\url{10.1063/5.0318344} (\bibinfo{year}{2026}).

\bibitem{rasmussen2005}
\bibinfo{author}{Rasmussen, C.~E.} \& \bibinfo{author}{Williams, C. K.~I.}
\newblock \emph{\bibinfo{title}{Gaussian {{Processes}} for {{Machine Learning}}}} (\bibinfo{publisher}{The MIT Press}, \bibinfo{year}{2005}).

\bibitem{powell1977}
\bibinfo{author}{Powell, M. J.~D.}
\newblock \bibinfo{journal}{\bibinfo{title}{Restart procedures for the conjugate gradient method}}.
\newblock {\emph{\JournalTitle{Mathematical Programming}}} \textbf{\bibinfo{volume}{12}}, \bibinfo{pages}{241--254}, \doiprefix\url{10.1007/BF01593790} (\bibinfo{year}{1977}).

\bibitem{auer2002}
\bibinfo{author}{Auer, P.}, \bibinfo{author}{{Cesa-Bianchi}, N.} \& \bibinfo{author}{Fischer, P.}
\newblock \bibinfo{journal}{\bibinfo{title}{Finite-time {{Analysis}} of the {{Multiarmed Bandit Problem}}}}.
\newblock {\emph{\JournalTitle{Machine Learning}}} \textbf{\bibinfo{volume}{47}}, \bibinfo{pages}{235--256}, \doiprefix\url{10.1023/A:1013689704352} (\bibinfo{year}{2002}).

\bibitem{wilson2017}
\bibinfo{author}{Wilson, J.~T.}, \bibinfo{author}{Moriconi, R.}, \bibinfo{author}{Hutter, F.} \& \bibinfo{author}{Deisenroth, M.~P.}
\newblock \bibinfo{journal}{\bibinfo{title}{The reparameterization trick for acquisition functions}}.
\newblock {\emph{\JournalTitle{arXiv:1712.00424}}} \doiprefix\url{10.48550/arXiv.1712.00424} (\bibinfo{year}{2017}).
\newblock \eprint{1712.00424}.

\bibitem{clarke1994}
\bibinfo{author}{Clarke, L.}, \bibinfo{author}{Glendinning, I.} \& \bibinfo{author}{Hempel, R.}
\newblock \bibinfo{title}{The {{MPI Message Passing Interface Standard}}}.
\newblock In \bibinfo{editor}{Decker, K.~M.} \& \bibinfo{editor}{Rehmann, R.~M.} (eds.) \emph{\bibinfo{booktitle}{Programming {{Environments}} for {{Massively Parallel Distributed Systems}}}}, \bibinfo{pages}{213--218}, \doiprefix\url{10.1007/978-3-0348-8534-8_21} (\bibinfo{publisher}{Birkh\"auser}, \bibinfo{address}{Basel}, \bibinfo{year}{1994}).

\bibitem{sobol1967}
\bibinfo{author}{Sobol', I.~M.}
\newblock \bibinfo{journal}{\bibinfo{title}{On the distribution of points in a cube and the approximate evaluation of integrals}}.
\newblock {\emph{\JournalTitle{USSR Computational Mathematics and Mathematical Physics}}} \textbf{\bibinfo{volume}{7}}, \bibinfo{pages}{86--112}, \doiprefix\url{10.1016/0041-5553(67)90144-9} (\bibinfo{year}{1967}).

\end{thebibliography}
\end{document}